\documentclass[english,11pt]{article}
\usepackage[utf8]{inputenc}
\usepackage{epsf,amsmath,amssymb,graphicx,dcolumn}
\usepackage{caption}
\usepackage{subcaption}
\usepackage{scalefnt,ulem,pstricks}
\usepackage{cite,hyperref}
\usepackage{cleveref}
\usepackage{todonotes}
\usepackage{color}
\usepackage{rotating}
\usepackage{microtype}
\definecolor{lightblue}{rgb}{.7,.8,1}
\Crefname{figure}{Fig.}{Figs.}
\title{\vspace*{-6em}
  \begin{flushright}
    {\sf\small
      ZU-TH 30/14 
      --- WUB/14-09
      --- LPN14-107 
    }
  \end{flushright}
\vspace*{2em} Finite top-mass effects in gluon-induced Higgs production with a jet-veto at \nnlo{}}
\author{Tobias Neumann$^a$ and Marius Wiesemann$^{b}$\\[2em]
$^a$ {\it Fachbereich C,
  Bergische Universit\"at Wuppertal,}\\[-.3em]
  {\it 42097 Wuppertal, Germany}\\
$^b$ {\it Physik-Institut, Universit\"at Z\"urich,}\\[-.3em]
 {\it 8057 Z\"urich, Switzerland}\\
{\small\tt tobias.neumann@uni-wuppertal.de}\\[-.3em]
{\small\tt marius.wiesemann@cern.ch}}

\date{}

\def\bal#1\eal{\begin{align}#1\end{align}}

\newcommand{\abbrev}{\scalefont{.9}}

\newcommand{\mhiggs}{m_{H}}

\newcommand{\mtop}{m_t}

\newcommand{\lhc}{{\abbrev LHC}}

\newcommand{\bsm}{{\abbrev BSM}}

\newcommand{\cp}{{\abbrev CP}}
\newcommand{\eft}{{\abbrev EFT}}

\newcommand{\pt}{\ensuremath{p_T}}
\newcommand{\ptjet}{\ensuremath{p_T^{\text{jet}}}}
\newcommand{\ptjetone}{\ensuremath{p_{T,1}^{\text{jet}}}}

\newcommand{\yjetone}{\ensuremath{y_{1}^{\text{jet}}}}
\newcommand{\ptveto}{\ensuremath{p_{T,\text{veto}}^{\text{jet}}}}
\newcommand{\ptmin}{\ensuremath{p_{T,\text{min}}^{\text{jet}}}}
\newcommand{\ptmax}{\ensuremath{p_{T,\text{max}}^{\text{jet}}}}

\newcommand{\nll}{\text{\abbrev NLL}}
\newcommand{\nnll}{\text{\abbrev NNLL}}
\newcommand{\lo}{\text{\abbrev LO}}
\newcommand{\nlo}{\text{\abbrev NLO}}
\newcommand{\nnlo}{\text{\abbrev NNLO}}

\newcommand{\sm}{{\abbrev SM}}
\newcommand{\qcd}{{\abbrev QCD}}

\newcommand{\plus}{{\abbrev +}}
\newcommand{\citere}[1]{Ref.\cite{#1}}
\newcommand{\citeres}[1]{Refs.\cite{#1}}
\newcommand{\eqn}[1]{Eq.\,(\ref{#1})}

\newcommand{\fig}[1]{Fig.\,\ref{#1}}

\newcommand{\sct}[1]{Section~\ref{#1}}

\newcommand{\app}[1]{\ref{#1}}
\newcommand{\dd}{{\rm d}}

\newcommand{\als}{\ensuremath{\alpha_s}}

\newcommand{\muF}{\mu_{\rm F}}
\newcommand{\muR}{\mu_{\rm R}}

\newcommand{\pdf}{{\abbrev PDF}}
\newcommand{\mstw}{{\abbrev MSTW2008} $68${\abbrev \%CL}}

\newcommand{\bld}[1]{\boldmath{$#1$}}

\begin{document}
\maketitle

\begin{abstract}

Effects from a finite top quark mass on the H+$n$-jet cross section through
gluon fusion are studied for $n=0/n\ge 1$ at \nnlo{}/\nlo{} \qcd{}. For this
purpose, sub-leading terms in $1/\mtop$ are calculated.  We show that the
asymptotic expansion of the jet-vetoed cross section at \nnlo{} is very
well behaved and that the heavy-top approximation is valid at the five permille
level up to jet-veto cuts of $300$\,GeV. For the inclusive Higgs+jet rate, 
we introduce a matching procedure that allows for a reliable prediction of the top-mass 
effects using the expansion in $1/\mtop$. The quality of the effective field
theory to evaluate differential K-factors for the distribution of the hardest jet
is found to be better than $1$-$2$\% as long as the transverse momentum of the
jet is integrated out or remains below about $150$\,GeV.
\end{abstract}

\section{Introduction}

The discovery of a scalar particle \cite{Aad:2012tfa,Chatrchyan:2012ufa} whose
properties are compatible with the particle causing the electro-weak symmetry
breaking predicted by the Standard Model (\sm{}), i.\,e. the Higgs boson, was
the first observation of a new elementary particle at the Large Hadron Collider
(\lhc{}). Initially, the discovery was based on the combination of various
experimental
search channels. By now, sufficient significance has been reached to claim an
observation alone in the two channels $H\rightarrow\gamma\gamma$
\cite{Aad:2013wqa,Khachatryan:2014ira} and $H\rightarrow ZZ^*\rightarrow 4l$
\cite{Aad:2013wqa,Chatrchyan:2013mxa}. Some of the experimental signatures rely
heavily on the analysis of particular phase-space regions of the final
state particles to reduce the contamination from the background processes. In
particular, in the search for $H\rightarrow WW^*\rightarrow l\nu l\nu$ 
\cite{Aad:2012me,Chatrchyan:2013iaa} the huge
\qcd{} background is reduced using a veto cut ($\ptjet<\ptveto$) on jets with a
large transverse momentum (\pt{}). The so-called jet-vetoed cross section
is used to lower specifically the $t\bar{t}$ and $tW$ background, where the
top quark mainly decays to high-\pt{} bottom quarks.

In the \sm{}, Higgs production proceeds predominantly through gluon fusion.\footnote{
The gluon fusion process has been studied in great detail over the
past years, see \citeres{Dittmaier:2011ti,Dittmaier:2012vm,Heinemeyer:2013tqa} 
and references therein.}
The jet-vetoed cross section in that case has been known up to \nnlo{} for a while
\cite{Catani:2001cr}. The residual uncertainties associated with this
observable have been subject to recent discussion \cite{Stewart:2011cf}, where
the resummation of logarithms in $\ptveto$ finally allowed to control these
uncertainties
\cite{Berger:2010xi,Banfi:2012yh,Tackmann:2012bt,Banfi:2012jm,Becher:2012qa,Liu:2012sz,Becher:2013xia,Stewart:2013faa}.
Another uncertainty, which is very specific to hadronic Higgs production
through gluon fusion, is induced by employing an effective theory approach,
where the top quark is assumed to be infinitely heavy, to determine higher
order corrections to the jet-vetoed rate. Recently, the full top- ($\mtop$) and
bottom-mass ($m_b$) dependence at \nlo{} has been added to the resummed
\nnlo{}\plus{}\nnll{} jet-veto efficiencies
\cite{Banfi:2013eda}.\footnote{Similarly, the full top- and bottom-mass effects
    on the \pt{} spectrum of the Higgs at \nlo{}\plus\nll{} have been
    considered in \citeres{Bagnaschi:2011tu,Mantler:2012bj,Grazzini:2013mca}.}
    At \nnlo{}, finite top-mass effects have been studied in case of the total
    cross section \cite{Harlander:2002wh,Anastasiou:2002yz,Ravindran:2003um} so
    far, which have been found to be below $\sim 1$\%
    \cite{Harlander:2009mq,Harlander:2009my,Pak:2009dg,Pak:2011hs}.
    Differential studies on the validity of the effective field theory approach
    at this order in the strong coupling constant (\als{}) have been considered
    only for Higgs quantities
    \cite{Harlander:2012hf}, but not for jet observables.\footnote{Further results are only available at
        leading order (\lo{}) in perturbation theory, for
        Higgs+$n$-jet production with $n=0,1,2$
        \cite{DelDuca:2001fn,DelDuca:2001eu,DelDuca:2003ba,Alwall:2011cy}.}
        They were found to be below $3$\% as long as the transverse momentum of
        the Higgs is integrated out or is below $\sim 150$\,GeV.

The goal of this paper is to validate the heavy-top approximation for  the
Higgs production cross section with a jet-veto at \nnlo{}. For this purpose, we
determine the expansion with respect to $1/\mtop^k$, where the leading term of
this series ($k=0$) corresponds to the effective field theory. Additionally, we
take into account the first and second non-trivial sub-leading term in the
$1/\mtop^k$ expansion ($k=2/4$). We supplement our analysis by further
jet-related quantities at \nlo{} such as the inclusive one-jet rate and
kinematical distributions of the hardest jet. For the jet-vetoed rate, we find
that finite top-quark effects for realistic experimental values of the jet-veto cut
($\ptveto\sim 30$\,GeV) are numerically negligible (about five permille). Even
for jet-veto cuts up to $600$\,GeV they remain below two percent. Therefore, we
conclude that the use of the effective field theory approach for the jet-vetoed
rate is fully justified.

This paper is organized as follows: In \sct{sec:outline}, we define the
jet-vetoed cross section and set-up the main ingredients of our
calculation. \sct{sec:results} contains our results, including our default
choices of the input parameters, some considerations at lower order and our
analysis of finite top-mass effects on the Higgs+$n$-jet cross section for
$n=0/n\ge 1$ at $\nnlo{}/\nlo{}$ as well as the \nlo{} \pt{} and rapidity ($y$)
distribution of the hardest jet. We conclude in \sct{sec:conclusions}.

\clearpage\section{Outline of the calculation}\label{sec:outline}

\begin{figure}
  \begin{center}
    \begin{tabular}{ccc}
      \mbox{\includegraphics[height=.12\textheight]{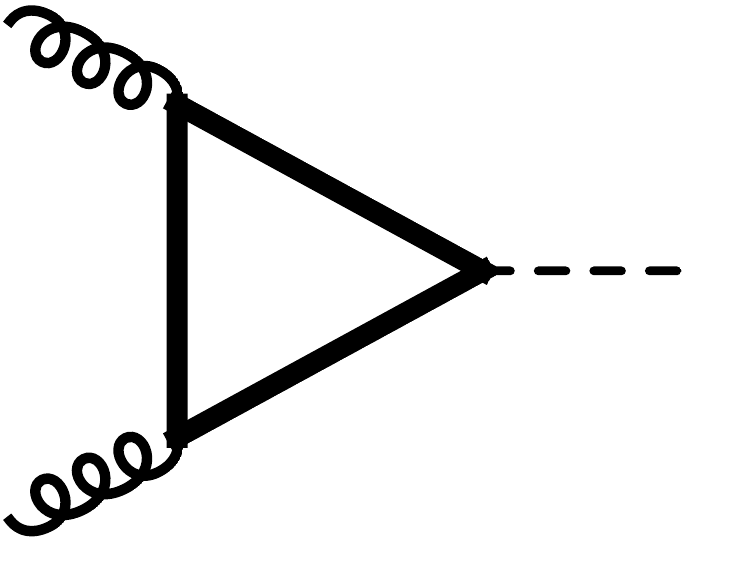}} &
      \mbox{\includegraphics[height=.12\textheight]{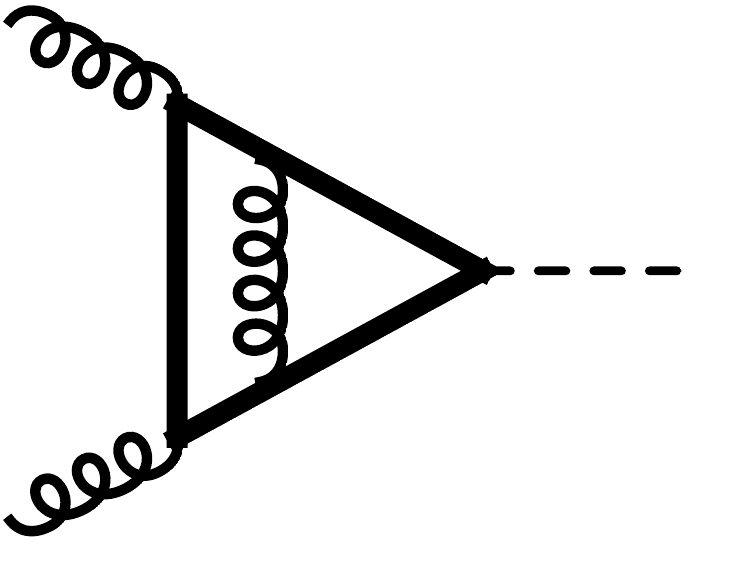}} &
      \mbox{\includegraphics[height=.12\textheight]{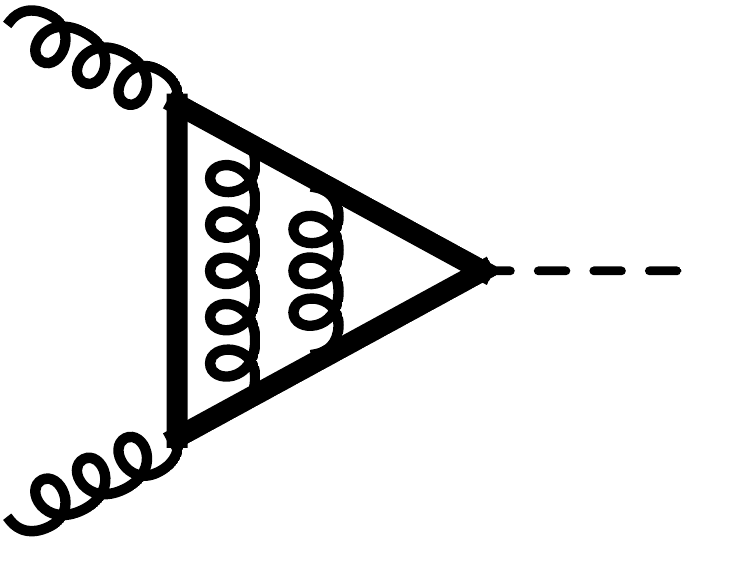}}
      \\[-0.15cm]
      (a) & (b) & (c)
    \end{tabular}\\[0.3cm]
    \parbox{.9\textwidth}{
      \caption[]{\label{fig:LOvirt}\sloppy{}A sample of Feynman diagrams
      contributing at $\pt = 0$. (a) \lo{} (one-loop); (b) \nlo{} (2-loop);
      (c) \nnlo{} (3-loop).  The graphical notation for the lines is: thick
      straight $\hat=$ top quark; spiraled $\hat=$ gluon; dashed $\hat=$ Higgs
      boson.
}}
  \end{center}
\end{figure}

\begin{figure}
  \begin{center}
    \begin{tabular}{ccc}
      \mbox{\includegraphics[height=.12\textheight]{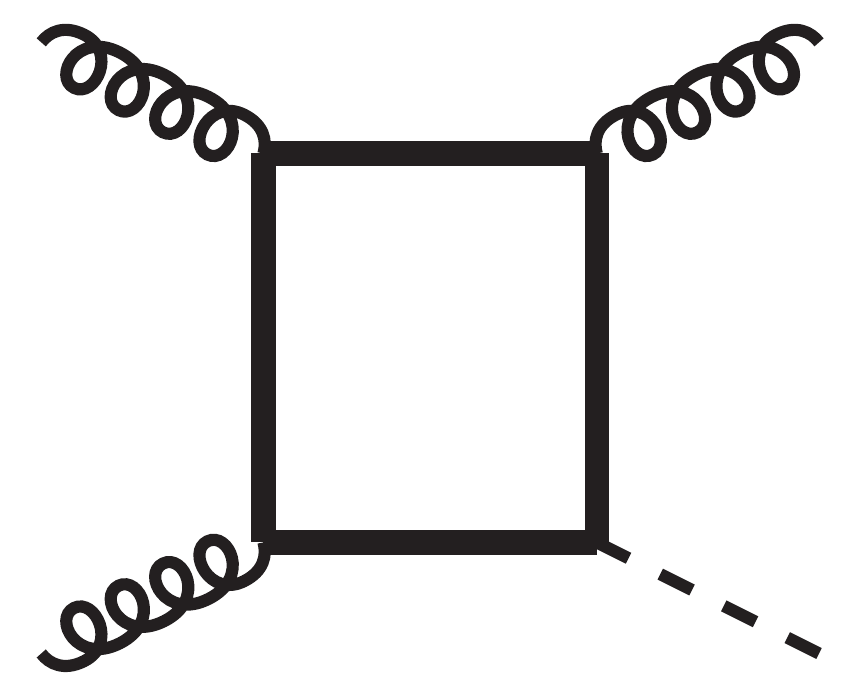}} &
      \mbox{\includegraphics[height=.12\textheight]{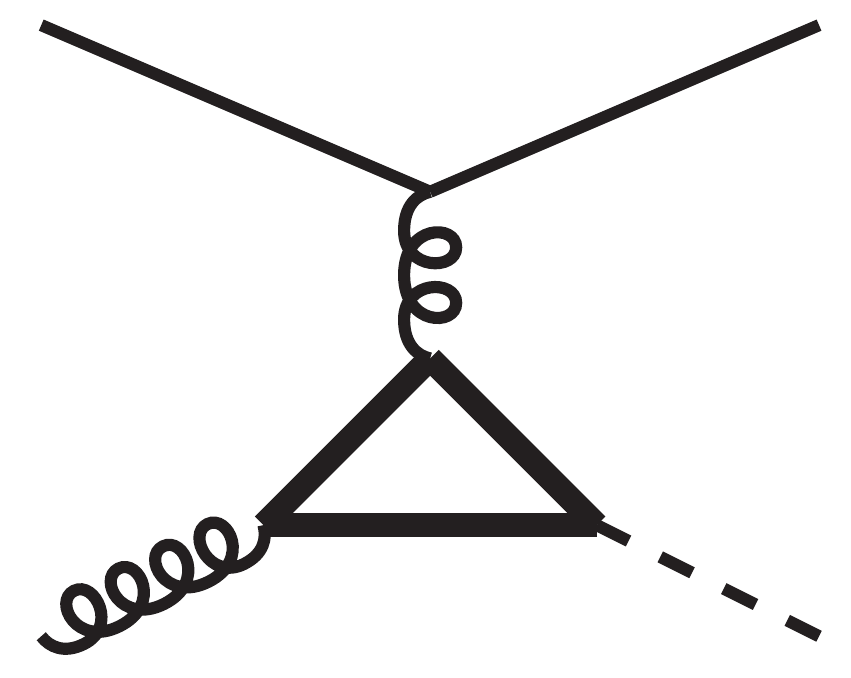}} &
      \raisebox{.7em}{\includegraphics[height=.12\textheight]{%
          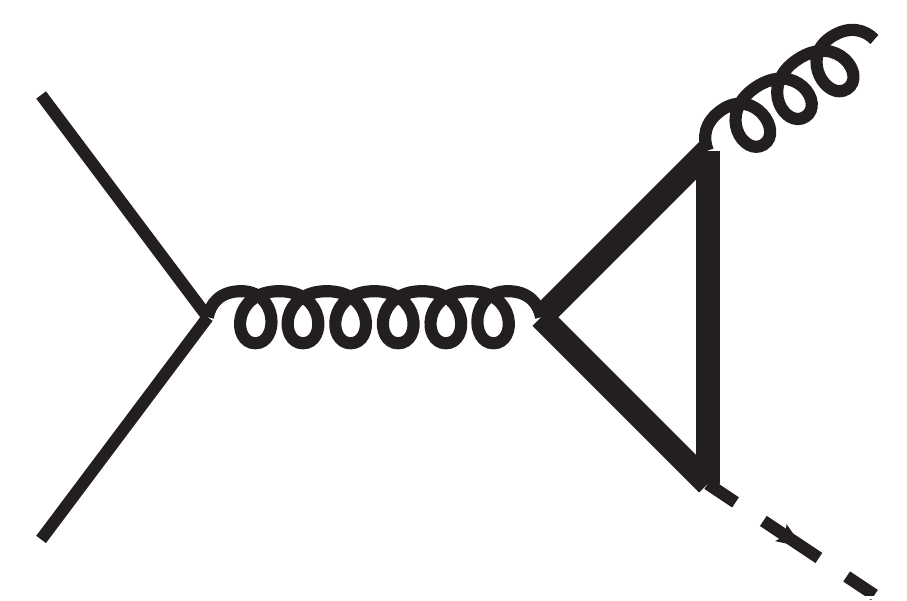}}
      \\[-0.15cm]
      (a) & (b) & (c)
    \end{tabular}\\[0.3cm]
        \begin{tabular}{cc}
      \mbox{\includegraphics[height=.12\textheight]{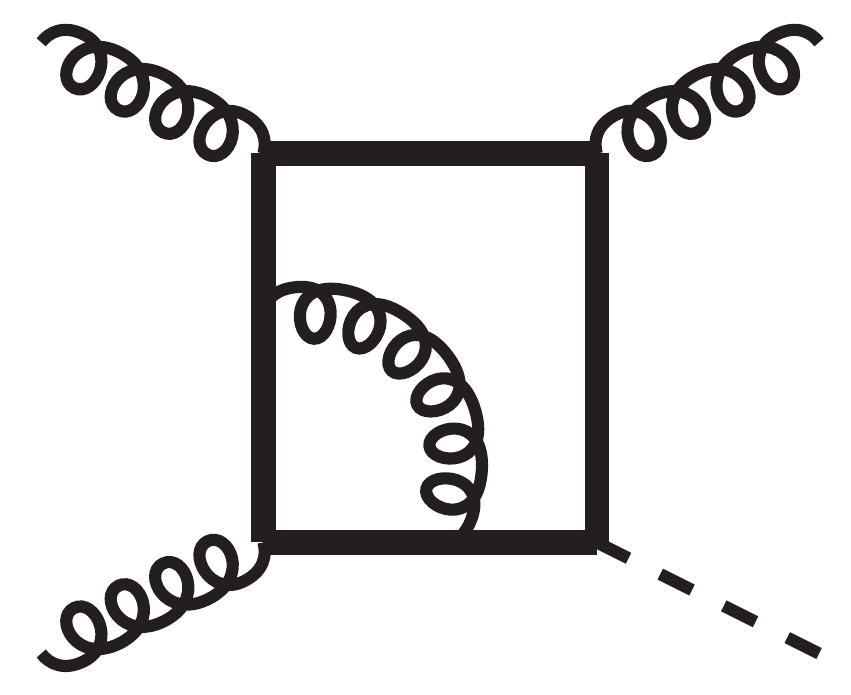}} &
      \raisebox{.2em}{\includegraphics[height=.12\textheight]{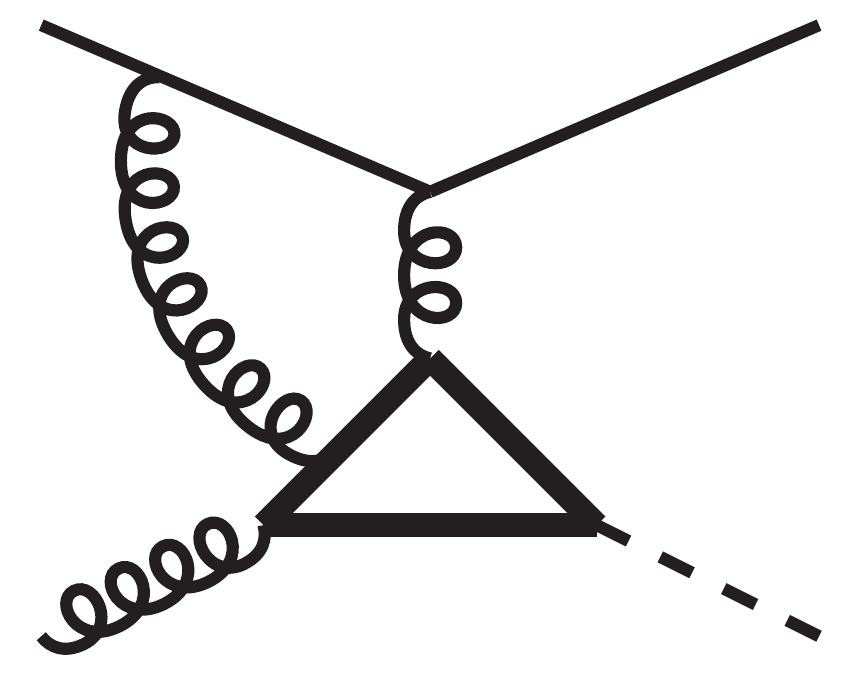}} 
      \\[-0.15cm]
      (e) & (f)
    \end{tabular}\\[0.3cm]
    \parbox{.9\textwidth}{
      \caption[]{\label{fig:real}\sloppy{}A sample of Feynman diagrams
      contributing at $\pt >0$. (a-c) single-real; (e-f) mixed real-virtual.
      The graphical notation for the lines is: thick straight $\hat=$ top
      quark; thin straight $\hat=$ light quark $q\in\{u,d,c,s,b\}$; spiraled
      $\hat=$ gluon; dashed $\hat=$ Higgs boson.
} }
  \end{center}
\end{figure}

\begin{figure}
  \begin{center}
    \begin{tabular}{cccc}
      \mbox{\includegraphics[height=.12\textheight]{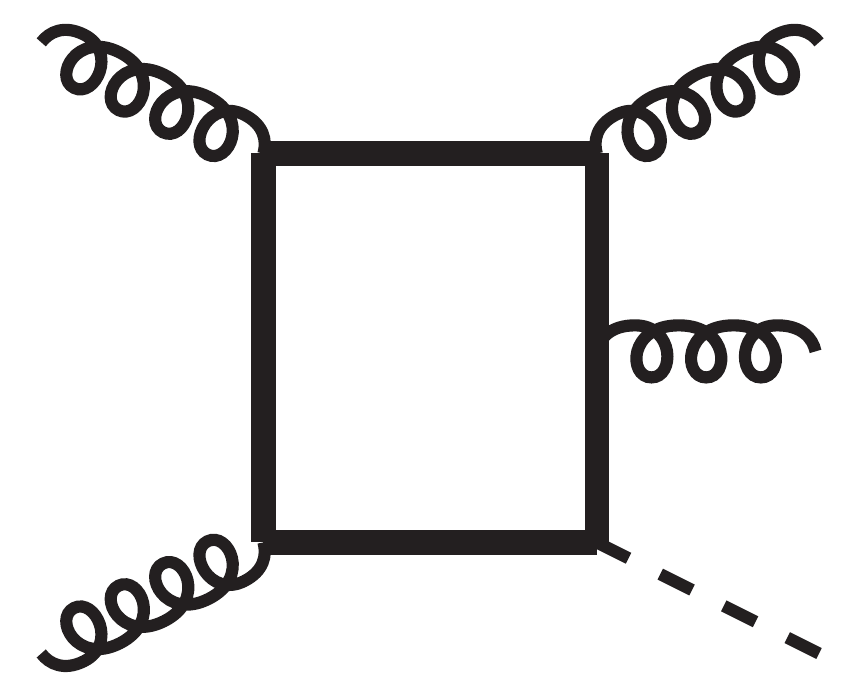}} &
      \mbox{\includegraphics[height=.12\textheight]{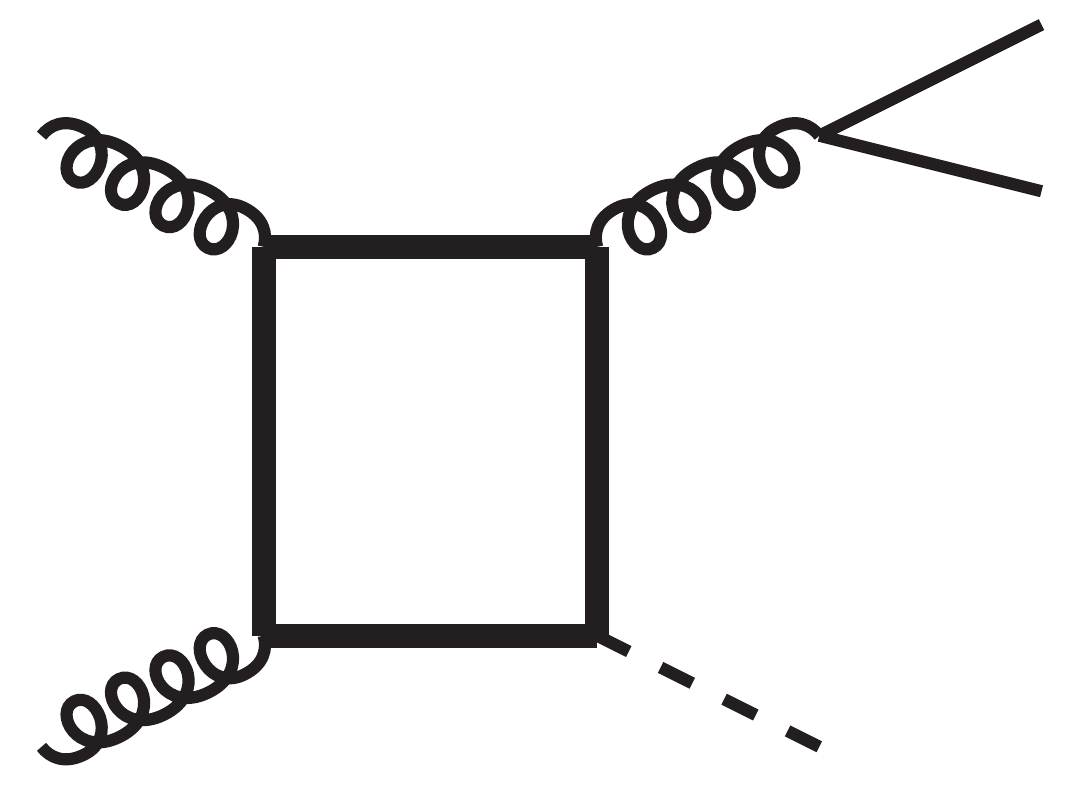}} &
      \mbox{\includegraphics[height=.12\textheight]{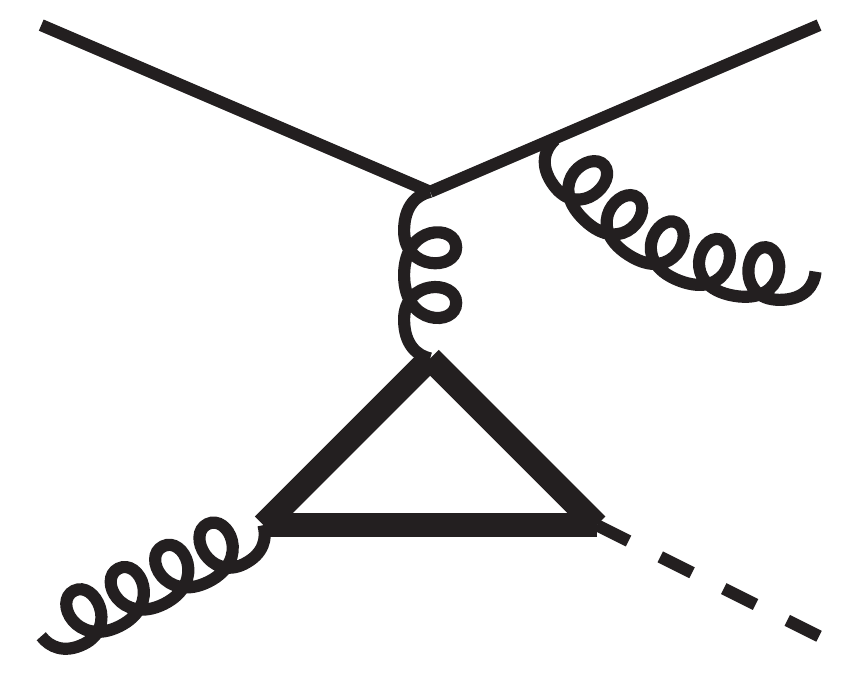}} &
      \mbox{\includegraphics[height=.12\textheight]{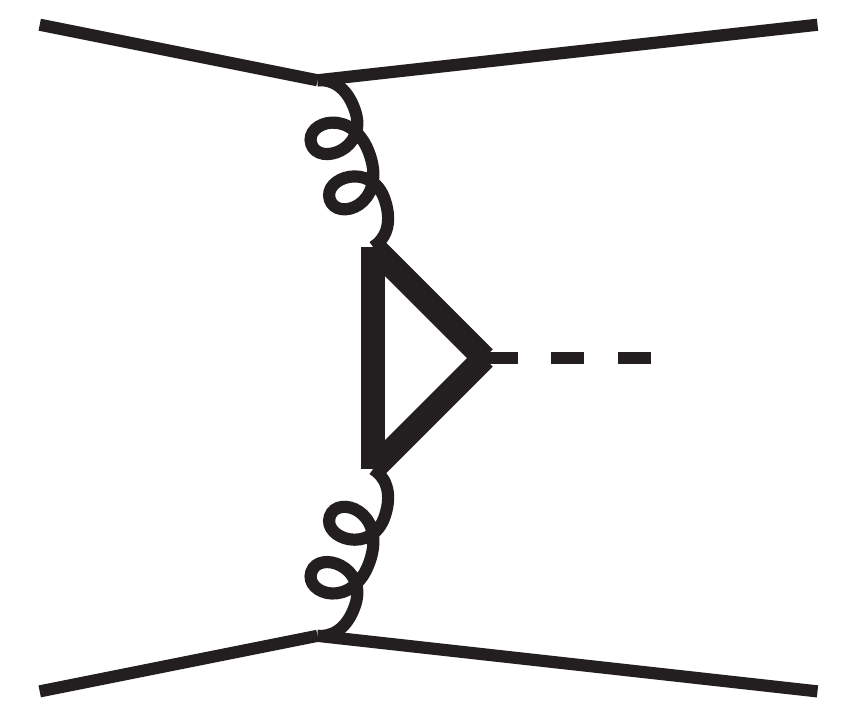}}
      \\[-0.15cm]
      (a) & (b) & (c) & (d)
    \end{tabular}\\[0.3cm]
    \parbox{.9\textwidth}{%
      \caption[]{\label{fig:real2}\sloppy
        Same as \fig{fig:real} but double real emission diagrams. } }
  \end{center}
\end{figure}

Considering the jet-vetoed (or $0$-jet) rate for Higgs production through gluon
fusion at \nnlo{}, various contributions have to be taken into account. At
\lo{}, the cross section is identical to the total rate, since the only
partonic process $gg\rightarrow H$ has no final state jets\footnote{Since we do
not include any parton showering or hadronization, ''jet`` denotes a cluster
of the outgoing partons throughout this paper.}, see \fig{fig:LOvirt}\,(a).
\fig{fig:LOvirt}\,(b) and (c) show two representative purely virtual diagrams
to $gg\rightarrow H$ entering at \nlo{} and \nnlo{}, respectively. The partonic
processes $gg\rightarrow H g$, $gq\rightarrow H q$ and $q\bar{q}\rightarrow H
g$ ($q\in\{u,d,s,c,b\}$) at one- and two-loop determine the single real and
mixed real-virtual contributions, see \fig{fig:real}.\footnote{Note that
    already the \lo{} process is loop-induced. Thus, the single real emission
diagrams contain one loop as well.} Examples for double real emission diagrams
are shown in \fig{fig:real2}, the corresponding processes $gg\rightarrow ggH$,
$gg\rightarrow q\bar{q}H$, $gq\rightarrow gqH$, $q\bar{q}\rightarrow
q\bar{q}H$, $q\bar{q}\rightarrow ggH$, $qq\rightarrow qqH$, $qq'\rightarrow qq'H$ and $\bar{q}q'\rightarrow \bar{q}q'H$ ($q'\neq q$) enter the calculation of the jet-vetoed
cross section at \nnlo{}. It is understood that the charge conjugated processes
must be included as well.

The most complicated Feynman diagrams are of the two-loop box-type and
three-loop-triangle-type with massless and massive (mass $\mtop$) internal and
one massive external line (mass $\mhiggs$)\footnote{$\mhiggs$ denotes the mass of the Higgs.}, see
Figs.\,\ref{fig:real}\,(e) and Figs.\,\ref{fig:LOvirt}\,(c), for example.
Although not out of reach, the complexity of the corresponding integrals is too
high for an efficient numerical evaluation. Thus, the \nnlo{} corrections are
known only in the effective theory approach with an infinitely heavy top quark
(heavy-top limit). Deploying this approximation the corresponding Feynman
diagrams simplify to one and two-loop level without internal masses and with an
effective Higgs-gluon vertex, multiplied by a Wilson coefficient which can be
evaluated perturbatively
\cite{Chetyrkin:1997un,Kramer:1996iq,Chetyrkin:1997iv,Schroder:2005hy,Chetyrkin:2005ia}.

In this paper, we go beyond the heavy-top approximation and study the effects
of a finite top-quark mass on the jet-vetoed rate. Therefore, we consider the
expansion of the cross section with respect  to $1/\mtop$, whose leading term
is given by the effective theory approach. We use the amplitudes which were
calculated in \citere{Harlander:2009mq} by applying automated asymptotic
expansions \cite{Harlander:1997zb,Steinhauser:2000ry,Smirnov:2002pj}.

In practice, we obtain the jet-vetoed Higgs cross section by removing all jet
contributions $\sigma_{\ge 1\text{-jet}}$ from the total rate
$\sigma_{\text{tot}}$. At \nnlo{} this reads \bal
\sigma_\text{veto}^\nnlo{} \equiv
\sigma^\nnlo_{0\text{-jet}}=\sigma^\nnlo_{\text{tot}}
-\sigma^{\nlo'}_{\ge 1\text{-jet}}\,,
\label{eq:jetveto}
\eal
where we use the prime-notation of \citere{Harlander:2011fx} to distinguish
$\sigma^{\nlo'}_{\ge 1\text{-jet}}$ calculated with \nnlo{} parton density
functions (\pdf{}s) from the proper \nlo{} quantity. For the total rate we
deploy the program {\tt ggh@nnlo}
\cite{Harlander:2002wh,Harlander:2009mq,Harlander:2009my,gghnnlo} including the
asymptotic expansion of the amplitudes in $1/\mtop{}^k$ up to
$k=6$.\footnote{We would like to thank Robert Harlander for providing a
private version of his code.} The calculation of the one-jet inclusive cross
section $\sigma_{\ge 1\text{-jet}}$ was carried out using the program described
in \citere{Harlander:2012hf}, where we implemented the anti-$k_T$ jet-algorithm
\cite{Cacciari:2008gp} to identify \qcd{} jets.\footnote{Since at most two jets
can occur in our calculation, the anti-$k_T$ leads to the same results as the
$k_T$ and the Cambridge-Aachen algorithm.} Furthermore, we extended its
capabilities to include sub-leading top-mass effects up to $1/\mtop^4$. Of
course, our setup allows to calculate the exclusive Higgs$+n$-jet rates for
$n=1$ and $n=2$ as well, where we work at \nlo{} and \lo{} accuracy,
respectively. 

A number of checks have been performed on our results. While the \pt{}
distribution of the Higgs in the heavy-top limit was checked
\cite{Harlander:2012hf} against the fixed order part of the program {\tt
HqT}~\cite{Bozzi:2003jy,Bozzi:2005wk,deFlorian:2011xf}, we used the program
{\tt HNNLO}\,\cite{Catani:2007vq,Grazzini:2008tf,Grazzini:2013mca} for a
numerical comparison of the jet-vetoed rate. The agreement was found to be 
better than one percent. At each order in the $1/\mtop$
expansion, we explicitly verified the independence of the $0$-jet rate with
respect to the so-called $\alpha$-parameter \cite{Nagy:1998bb,Nagy:2003tz},
which allows to restrict the phase space of the Catani-Seymour dipoles
\cite{Catani1996}. The asymptotic expansion of the amplitudes as well as the
program {\tt ggh@nnlo} have been validated previously by the agreement of the
inclusive cross section between \citeres{Harlander:2009mq} and
\cite{Pak:2009dg}. 

As observed in \citeres{Harlander:2009my,Pak:2011hs,Harlander:2012hf},
the $1/\mtop$ expansion provides no valid description for the purely quark-induced channels.
Therefore, they constitute a solid, though rather minor limitation of the effective field theory,
since their contribution is more than two orders of magnitude smaller than the sum
of all channels. We can therefore safely disregard them from our considerations.

\section{Results}\label{sec:results}

\subsection{Input parameters}\label{sec:input}
We study finite top-mass effects on Higgs+$n$-jet cross sections for $n=0/n\ge
1$ in the gluon fusion process at the \lhc{} with 13\,TeV center-of-mass
energy. Our choice of the central factorization and renormalization scale is
$\muF=\muR=\mhiggs$. All numbers are produced with the \mstw{} \pdf{}s
\cite{Martin:2009iq} which implies that the numerical value for the strong
coupling constant is taken as $\als (m_Z ) = 0.13939$ at \lo{}, $\als (m_Z ) =
0.12018$ at \nlo, and $\als (m_Z ) = 0.11707$ at NNLO. We
set the on-shell top quark mass to $\mtop = 173.5$\,GeV.

Jets are defined using the anti-$k_T$ algorithm \cite{Cacciari:2008gp} with jet
radius: $R = 0.5$. Unless stated otherwise, a jet is required to have transverse
momentum of $\ptjet > 30$\,GeV, while we apply no cuts on the Higgs
momentum.\footnote{We checked that our results directly generalize
    to experimentally applied jet definitions for this process which usually
    imply $\ptjet>25$-$30$\,GeV and a rapidity cut \cite{Aad:2012me,Chatrchyan:2013iaa}.}

\subsection{Notation}
To deal with the additional expansion of the cross section with respect to
$1/\mtop$ we introduce the following notation: The truncation of the cross
section is defined by
\bal
\label{eq:defcs}
\left[\dd\sigma^{X}\right]_{1/\mtop^k},\qquad X\in\{\lo{},\nlo{},\nnlo{}\},\qquad k\in\{0,2,4,\ldots\}\,
\eal
where $X$ denotes the order of perturbation theory and $k$ the order at which
the $1/\mtop^k$ expansion of the cross section is truncated. If the index
$1/\mtop^k$ and the brackets are absent, it means that the cross section is not
truncated and, consequently, $\dd\sigma^{X}$ denotes the cross section with
exact top-mass dependence. Here and in what follows we imply that all cross sections 
are reweighted by the exact top-mass dependence at \lo{}:
\bal
\left[\dd\sigma^{X}\right]_{1/\mtop^k}\equiv \left[\dd\bar\sigma^{X}\right]_{1/\mtop^k}\cdot \sigma^{\lo{}}/\left[\sigma^{\lo{}}\right]_{1/\mtop^k},
\eal
where $\dd\bar\sigma$ denotes the unweighted cross section and $\sigma^{\lo{}}$ the Born-level cross section for $gg\rightarrow H$.

In order to analyse the perturbative corrections to the cross section, we
define the $K$-factor
\begin{equation} \begin{split} K^X_k(b)&=\frac{\left[\dd
        \bar\sigma^X(b)\right]_{1/\mtop^k}}{ \left[\dd
        \bar\sigma^\lo{}(b)\right]_{1/\mtop^k}}\,.  \label{eq:K} \end{split}
\end{equation}
On the right hand side of this definition, it is understood that $\dd
\sigma(b)$ is integrated over all kinematical variables {\it except} the set
$b$, where we consider $b=\{\ptjetone\}$ and $b=\{\yjetone\}$ (i.\,e., transverse momentum and rapidity distributions of the hardest jet).
For example, $K_0^\nlo{}(p_T^H)$ is the \nlo{} K-factor in the heavy-top limit
of the \pt{} distribution of the Higgs which has been found to be valid at the
$2$-$3$\% level for $p_T^H\lesssim 150$\,GeV
\cite{Harlander:2012hf}.\footnote{Note that in \app{app:pTH} we extend the
analysis of the transverse momentum distribution in \citere{Harlander:2012hf}
by considering an additional term in the $1/\mtop^k$ expansion ($k=4$).} Using
the $1/\mtop{}$ expansion, we will study whether this observation can be
expected to carry over also to jet quantities.

\subsection{Lower order results}
\label{sec:loword}
\begin{figure}
	\begin{subfigure}[b]{0.5\textwidth}
         	\includegraphics[width=\textwidth]{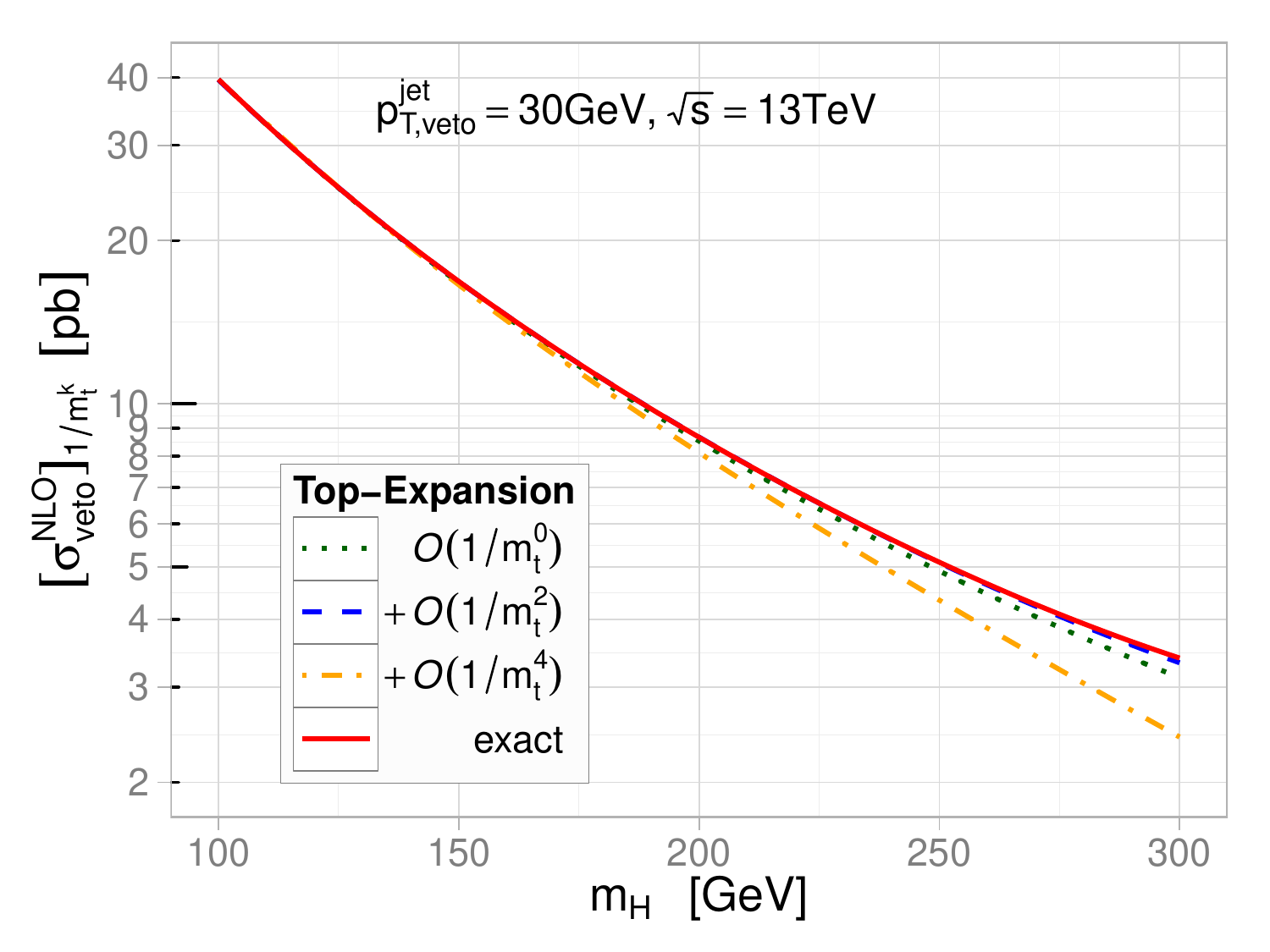}
	        \caption{}
            \label{fig:veto_NLO_mH_abs}
        \end{subfigure}%
	\begin{subfigure}[b]{0.5\textwidth}
         	\includegraphics[width=\textwidth]{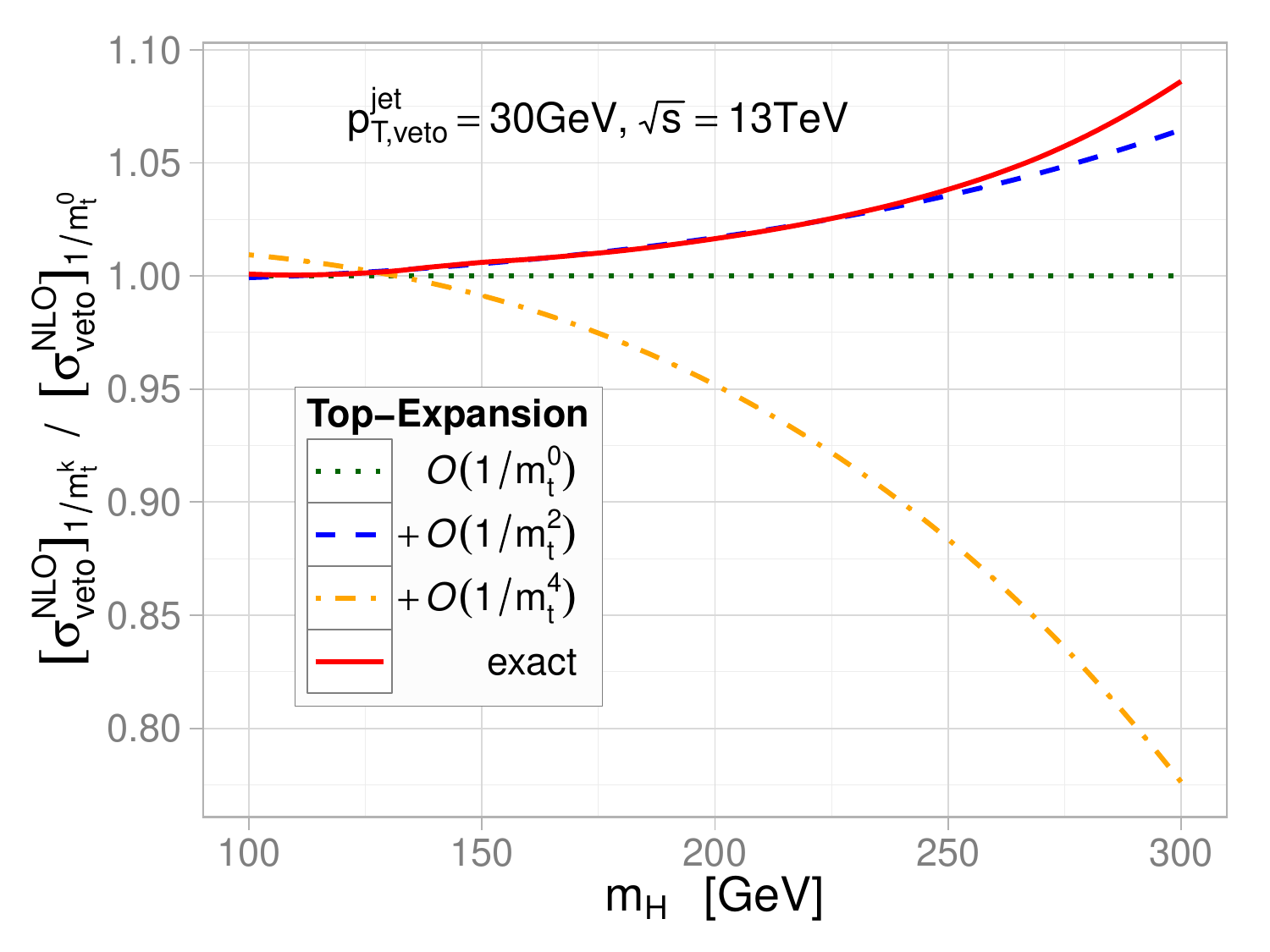}
		\caption{}
        \label{fig:veto_NLO_mH_rel}
        \end{subfigure}%
\begin{center}
        \parbox{.9\textwidth}{\caption{\label{fig:veto_NLO_mH}Higgs+$0$-jet
        cross section at \nlo{} including terms up to $1/\mtop^k$ as a function
        of $\mhiggs$ for $\ptveto=30$\,GeV. Dotted/dashed/dash-dotted: $k =
        0/2/4$. (a) absolute; (b) normalized to $k=0$.
        }} \end{center}
\end{figure}

\Cref{fig:veto_NLO_mH_abs} shows the \nlo{} jet-vetoed cross section as a
function of the Higgs mass. We applied a veto of $\ptveto=30$\,GeV
on the jet transverse momenta. At this order, the exact dependence on the
top-quark mass is known (solid curve).\footnote{To obtain the \nlo{} total cross section with exact top-mass dependence we employed the code {\tt SusHi} \cite{Harlander:2012pb}.} Comparing it to the expansion of the
cross section up to $1/\mtop^k$ for $k=0$ (dotted curve), $k=2$  (dashed curve)
and  $k=4$ (dash-dotted curve), we can assess the quality of the $1/\mtop$
expansion. Clearly, its convergence starts deteriorating once the Higgs mass
exceeds the top-quark mass.

In general, the aim of our analysis is to obtain accurate predictions including
mass effects for the various jet observable considered in this paper and to use them
to estimate the mass corrections with
respect to the effective field theory (\eft{}). The deviation of the higher
orders in the asymptotic expansion from the leading term indicates the validity
of the \eft{} to approximate the cross section in the full theory. For this
purpose, we normalize all curves to the $1/m_t^0$ approximation in
\Cref{fig:veto_NLO_mH_rel}. For small values of $\mhiggs{}$, the mass effects
are at the percent level. While the expansion up to $1/m_t^2$ remains extremely
close to the full result over the whole mass range, the $1/m_t^4$ corrections
reduce the cross section significantly towards larger values of $\mhiggs{}$.
Assuming the exact cross section was not known, which is the case at \nnlo{},
we would therefore estimate the uncertainty of the mass corrections on the
\eft{} result to be below $5\%$ for $\mhiggs\lesssim 200$\,GeV. Fortunately,
all orders of the $1/\mtop$ expansion coincide to a very good accuracy at
$m_H\simeq125\,\mathrm{GeV}$.

\begin{figure}
	\begin{subfigure}[b]{0.5\textwidth}
         	\includegraphics[width=\textwidth]{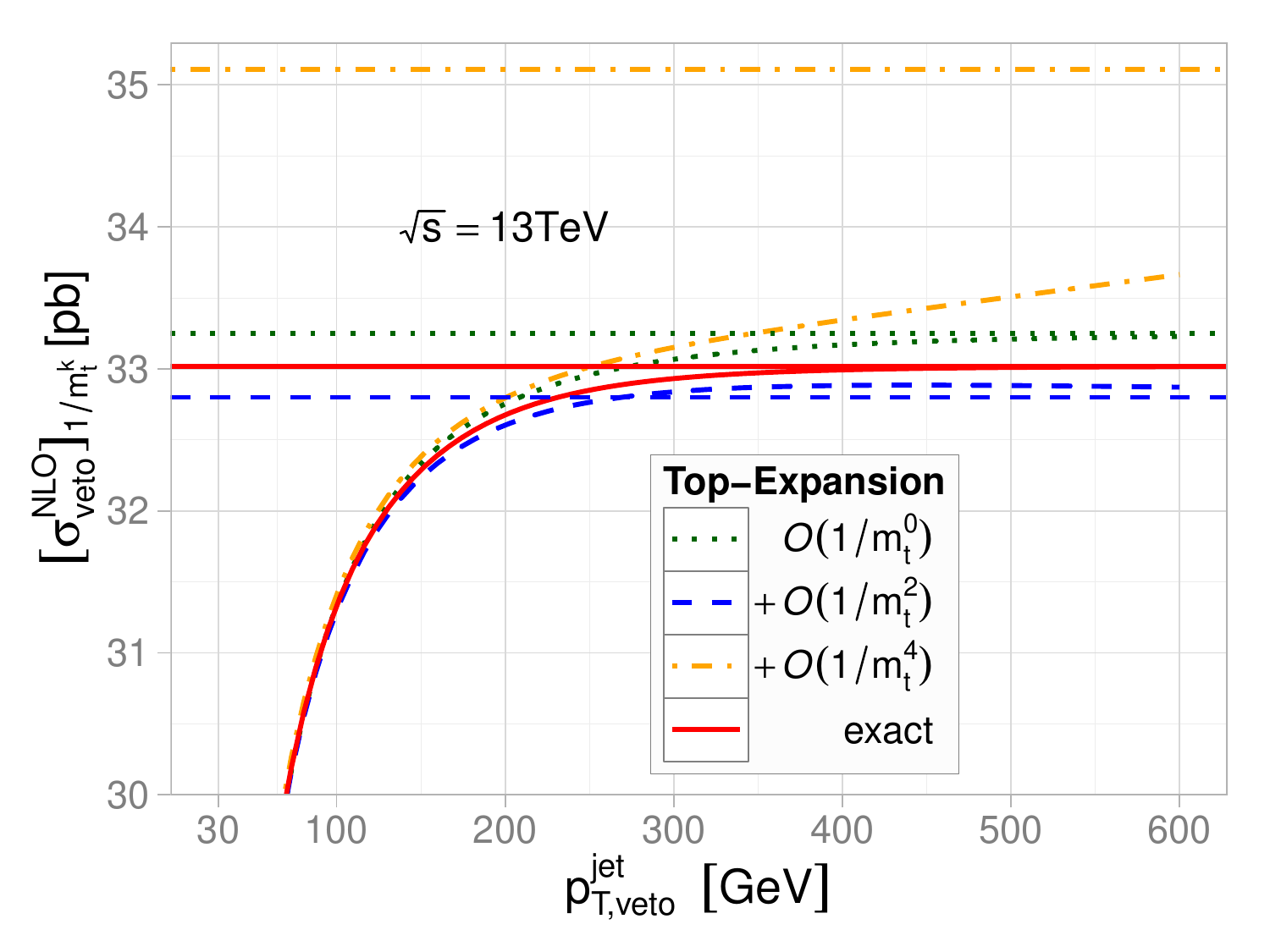}
	        \caption{}
            \label{fig:jetvetoNLO_abs}
        \end{subfigure}%
	\begin{subfigure}[b]{0.5\textwidth}
         	\includegraphics[width=\textwidth]{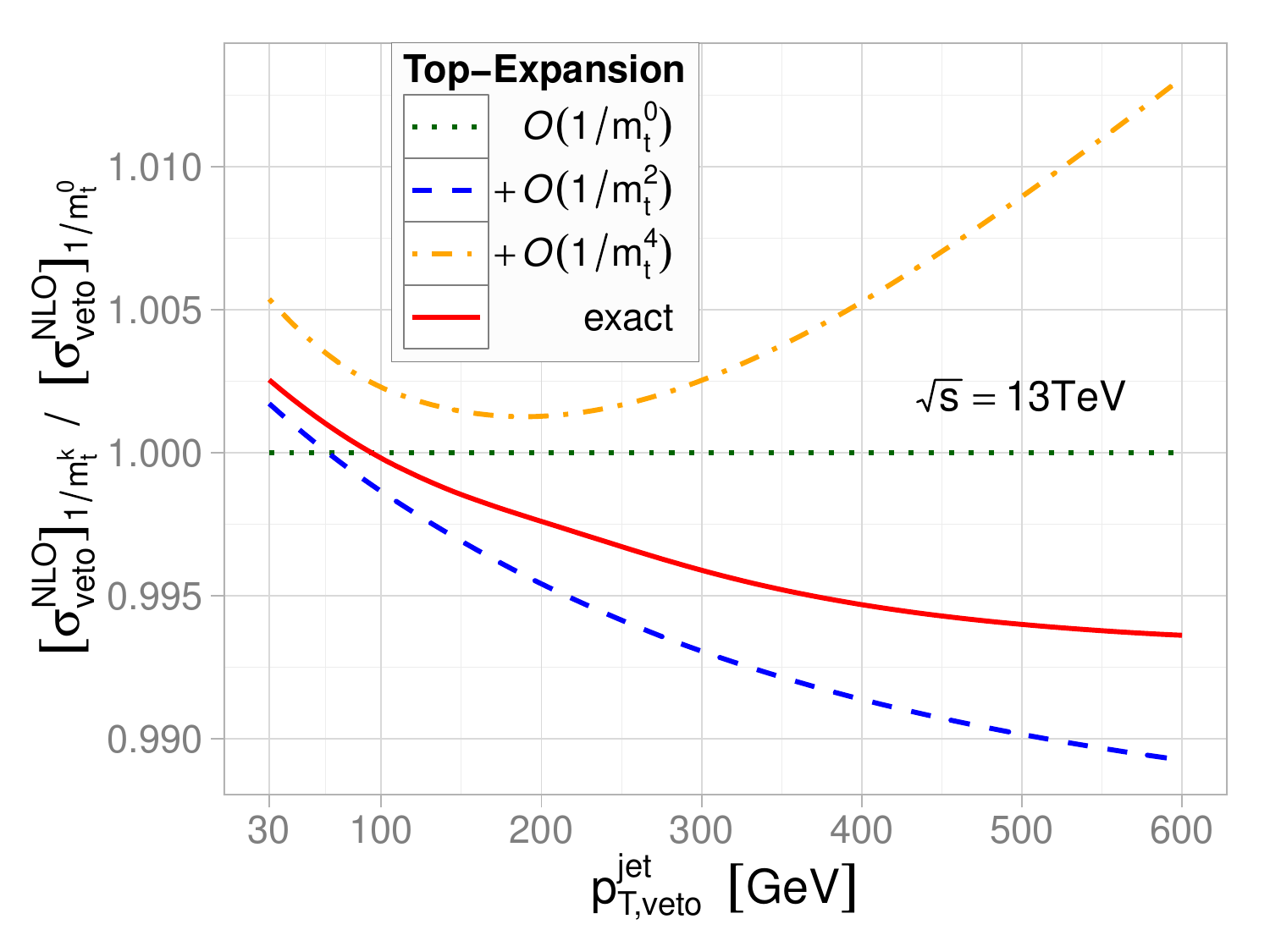}
		\caption{}
        \label{fig:jetvetoNLO_rel}
        \end{subfigure}%
\begin{center}
    	\parbox{.9\textwidth}{%
    \caption{\label{fig:jetvetoNLO}{Higgs+$0$-jet cross section at \nlo{}
    including terms up to $1/\mtop^k$ as a function of $\ptveto{}$.
    Dotted/dashed/dash-dotted: $k = 0/2/4$. (a) absolute; (b) normalized to
    $k=0$.
    }}}
\end{center}
\end{figure}

In \Cref{fig:jetvetoNLO_abs}, we study the top-mass corrections to the \nlo{} cross
section as a function of the jet-veto cut for $\mhiggs=125.6$\,GeV. The
horizontal lines denote the total inclusive cross sections, which correspond to
$\ptveto{}\rightarrow\infty$.  The agreement between the curves is remarkable.
While the differences are at the permille-level for small jet-veto cuts they
remain below $2.5$\% even at $\ptveto=600$\,GeV. Again, the asymptotic
expansion leads to a proper estimation of the mass effects, not underestimating
the uncertainty induced by the heavy-top approximation with respect to exact
one. Therefore, the $1/\mtop$ terms can be expected to yield a conservative
validation of the \eft{} as well at \nnlo{}.

The reason that the $1/\mtop$ expansion of the jet-vetoed rate is well behaved
even beyond the $2\,\mtop$ threshold is the phase-space suppression, which
strongly reduces contributions from hard jets. However, the $1/\mtop^4$ term
receives unjustified large contribution from $\ptjet\gtrsim 400$\,GeV. In that
region, $\sigma_{\text{veto}}^\nlo{}$,
$[\sigma_{\text{veto}}^\nlo{}]_{1/\mtop^0}$ as well as
$[\sigma_{\text{veto}}^\nlo{}]_{1/\mtop^2}$ develop a flat behavior, which is
expected from phase-space suppression, while
$[\sigma_{\text{veto}}^\nlo{}]_{1/\mtop^4}$ grows almost linearly.
This reveals
that the convergence of the amplitudes at $1/\mtop^4$ in the large-\pt{} tail
is broken.  The previous observations are in direct analogy to the total cross
section. In this case, the bulk of the cross section originates from the region
$\sqrt{s}\lesssim 2\,\mtop$, in which the asymptotic expansion is well behaved
\cite{Harlander:2009my}.  Nevertheless, the $1/\mtop^4$ term receives huge
contributions as $\sqrt{s}\gg 2\,\mtop$ \cite{Harlander:2009my}, since the
convergence of the amplitudes is spoiled at large energies. In fact, looking at
the total cross sections in \Cref{fig:jetvetoNLO_abs}, it is obvious that the
leading and first sub-leading term in the asymptotic expansion compare better
to the exact result than when including the $1/\mtop^4$ terms.\footnote{
Note that we applied no matching of the total inclusive cross section to the
high-energy limit here which will be discussed below.}

\begin{figure}
	\begin{subfigure}[b]{0.5\textwidth}
         	\includegraphics[width=\textwidth]{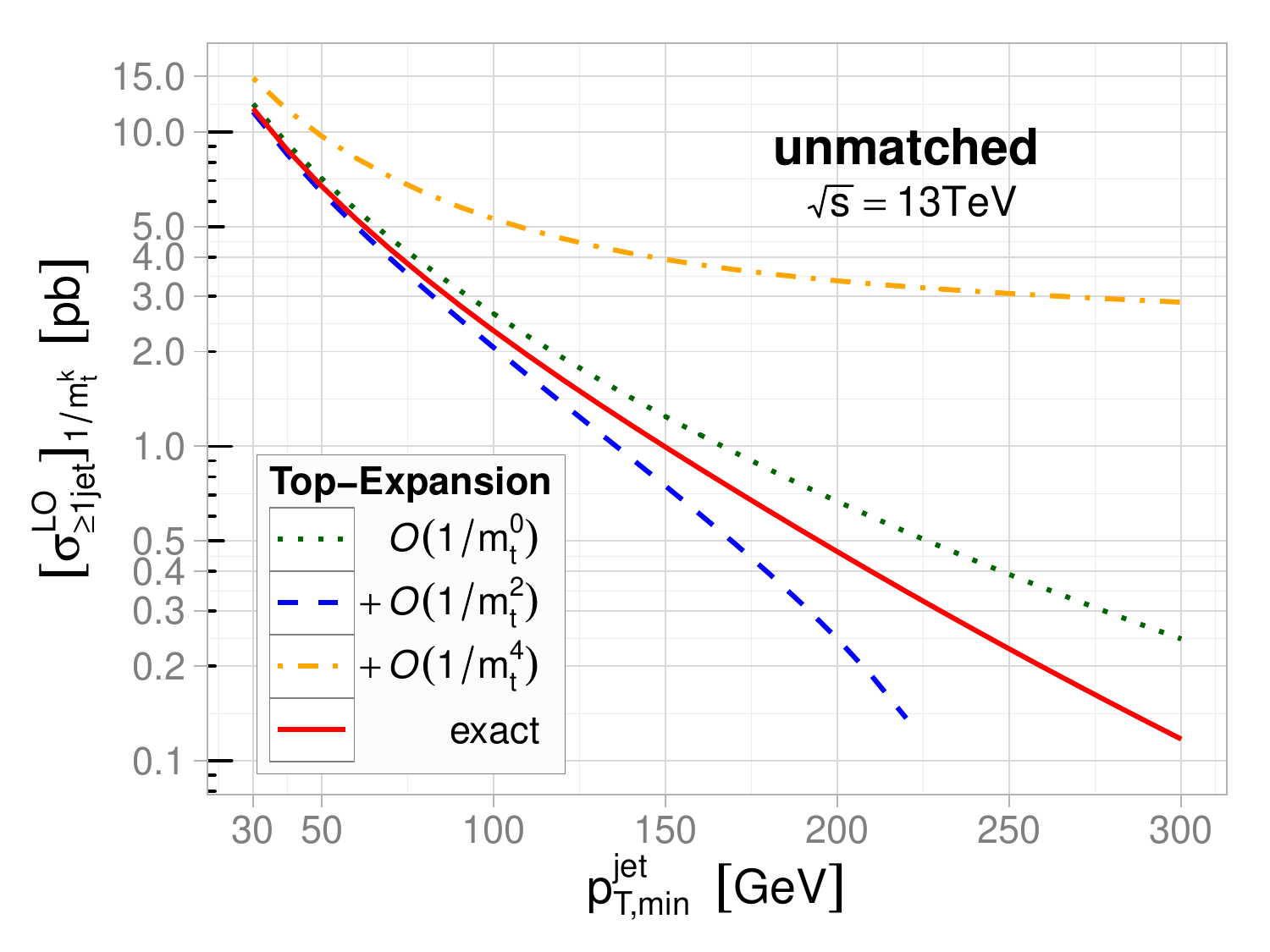}
	        \caption{}
            \label{fig:onejetLO_unmatched}
        \end{subfigure}%
	\begin{subfigure}[b]{0.5\textwidth}
         	\includegraphics[width=\textwidth]{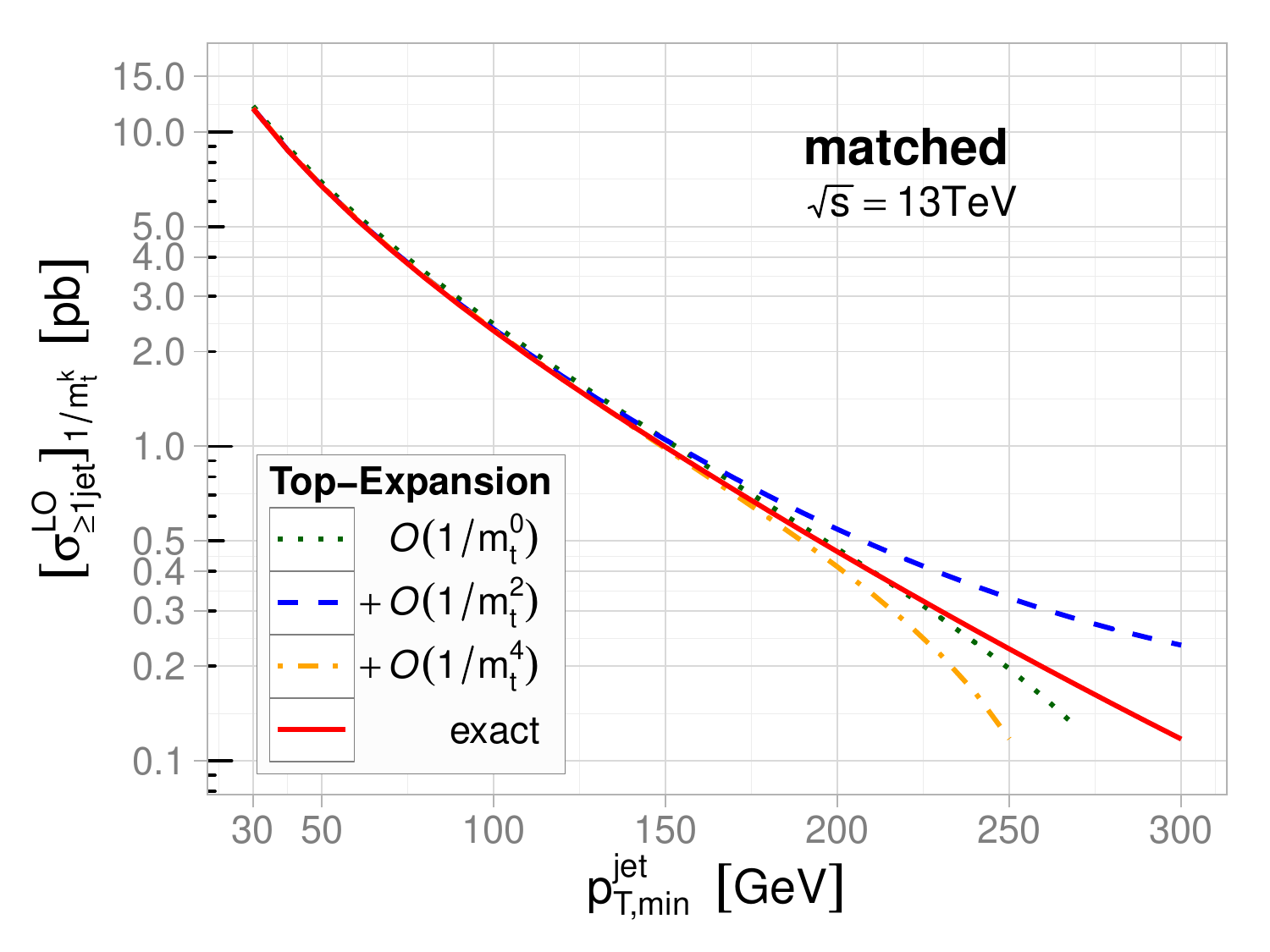}
		\caption{}
        \label{fig:onejetLO_matched}
        \end{subfigure}%
\begin{center}
    	\parbox{.9\textwidth}{%
    \caption{\label{fig:onejetLO}{Inclusive Higgs+jet cross section at \lo{}
    including terms up to $1/\mtop^k$ as a function of $\ptmin{}$.
    Dotted/dashed/dash-dotted: $k = 0/2/4$. (a) unmatched; (b) matched
    according to \eqn{eq:matchLO}.}}} \end{center}
\end{figure}

To obtain the inclusive Higgs+jet cross section a cut $\ptjet>\ptmin$ is
applied, which removes the bulk of the well behaved soft jets and, therefore,
enhances the contribution from the problematic large-\pt{} region.
\Cref{fig:onejetLO_unmatched} compares the $1/\mtop^k$ expansion of the
inclusive Higgs+jet rate at \lo{} for $k=0/2/4$ to the exact result. While
already at $\ptmin=30$\,GeV the deviation between the curves relative to
$1/\mtop^0$ is quite large ($\sim 27$\%), convergence of the asymptotic
expansion is completely lost at large values of $\ptmin$.  Thus, we cannot use
the ordinary $1/\mtop{}$ expansion to determine a sensible estimate of the mass
effects on the inclusive Higgs+jet rate.

However, the same problematic effects contribute to the total inclusive cross
section $\sigma_{\text{tot}}$, as we have seen before.  In this case, a
matching to the high-energy limit was performed as described in
\citere{Harlander:2009my} to control the region $\sqrt{s}>2\,\mtop$. Similarly,
a matching of the inclusive Higgs+jet cross section to the $\pt\rightarrow
\infty$ limit would temper unjustified effects from high-$\pt{}$ jets. Let us
assume this matched cross section was known and call it
$\sigma_{\ge 1\text{-jet,\,matched}}$.  Given the fact that the total cross
section can be viewed as the integral over the \pt{} distribution and the
asymptotic expansion in the small-\pt{} region works almost perfectly, the
following relation  
should be valid up to a very good precision as long as $\ptmin$ remains at
moderate values:\footnote{With "moderate values`` we mean values at which the
asymptotic expansion works well. The usual jet definitions with $\ptmin\sim
30$\,GeV are well within that region.}
\bal \label{eq:same}
\left[\sigma^\nlo_{\text{tot,\,matched}}\right]_{\mtop^k} -
\left[\sigma^\nlo_{\text{tot,\,unmatched}}\right]_{\mtop^k}  =
\left[\sigma^{\lo'}_{\ge 1\text{-jet,\,matched}}\right]_{\mtop^k}  -
\left[\sigma^{\lo'}_{\ge 1\text{-jet,\,unmatched}}\right]_{\mtop^k} \,,
\eal
where the primed \lo{} quantity is calculated with \nlo{} parton distributions,
as defined in \sct{sec:outline}. This equation allows us to determine the
matched inclusive Higgs+jet cross section by using \lo{} \pdf{}s for all
quantities:
\bal \label{eq:matchLO} \left[\sigma^{\lo}_{\ge
1\text{-jet,\,matched}}\right]_{\mtop^k} \equiv \left[\sigma^{\lo}_{\ge
1\text{-jet,\,unmatched}}\right]_{\mtop^k}  +
\left[\sigma^{\nlo^*}_{\text{tot,\,matched}}\right]_{\mtop^k}  -
\left[\sigma^{\nlo^*}_{\text{tot,\,unmatched}}\right]_{\mtop^k}  \,,
\eal
where we defined the starred \nlo{} cross section to be evaluated with \lo{}
\pdf{}s. \Cref{fig:onejetLO_matched} shows the matched cross section as defined
in \eqn{eq:matchLO}. It is very impressive how close all curves are to the
exact result with respect to the unmatched case in \Cref{fig:onejetLO_unmatched}. 

\begin{figure}
\begin{center}
    \includegraphics[width=0.65\textwidth]{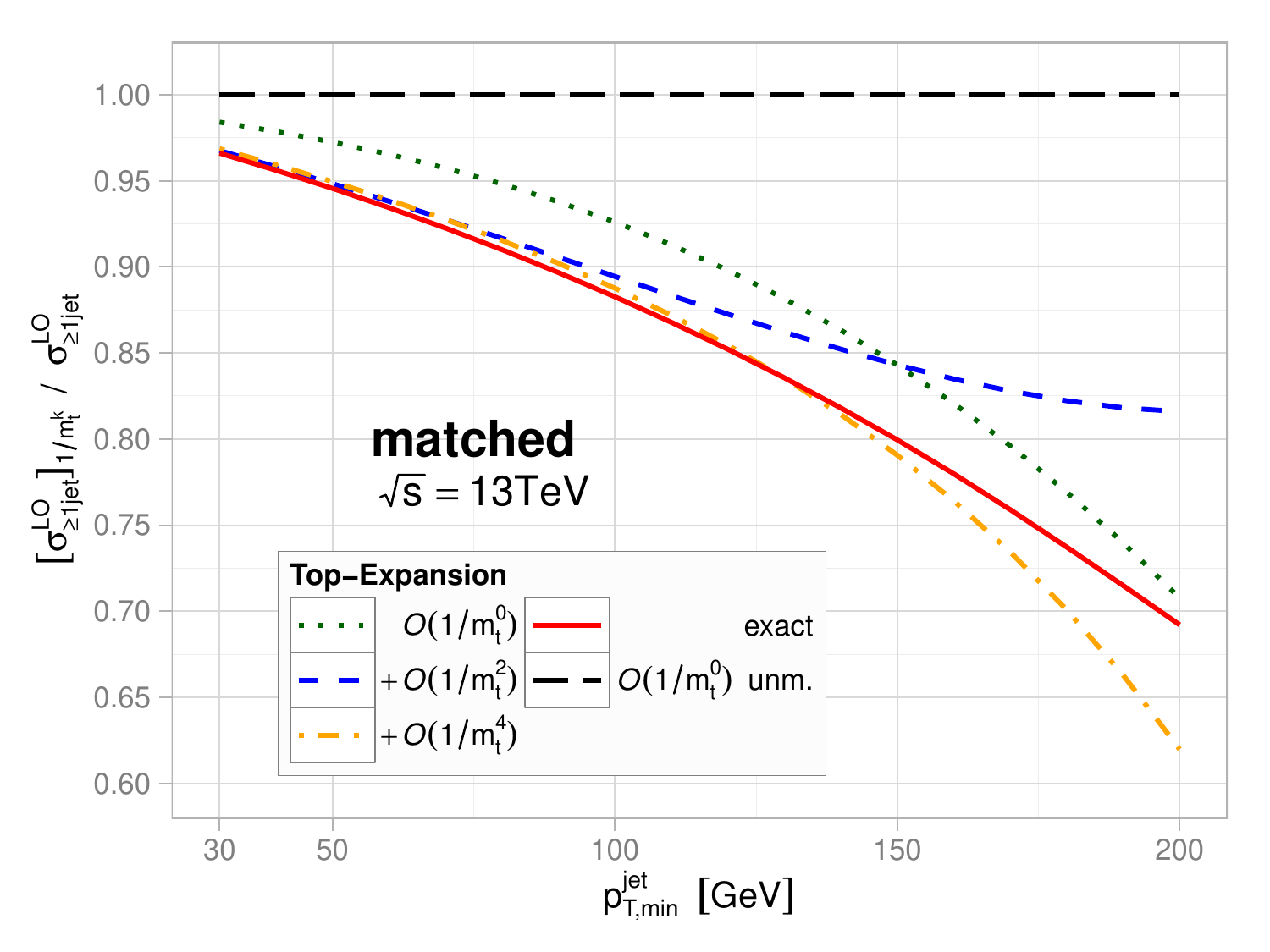}
	\parbox{.9\textwidth}{%
    \caption{\label{fig:onejetLOrel}Same as \Cref{fig:onejetLO_matched}, but
    normalized to unmatched $1/m_t^0$ cross section (dotted curve of
    \Cref{fig:onejetLO_unmatched}).}}
\end{center}
\end{figure}

In \Cref{fig:onejetLOrel}, the matched predictions of
\Cref{fig:onejetLO_matched} are normalized to unmatched cross section in the
heavy-top limit (dotted curve in \Cref{fig:onejetLO_unmatched}). Comparing first the
matched cross sections to the exact curve, their overall agreement is
remarkable ($\lesssim 5$\% for $\ptmin \le 150$\,GeV). In that region, they are
successively closer to the exact result, as $k$ increases. The deviation of
the \eft{} result from the matched curves on the other hand allows its
validation at the $3-10$\% level for $\ptveto \in[30,100]$\,GeV. Thus, with the
definition of the matched cross section we recovered the ability to validate
the heavy-top limit for the inclusive Higgs+jet rate. This will prove useful at
\nlo{}, where the exact result is not available.

There are cases in our analysis where the reliability of the $1/\mtop{}$
expansion appears to be exceptionally good. This
happens when the $1/m_t^4$ corrections become negligible and, consequently, the
expansions up to $1/m_t^2$ and up to $1/m_t^4$ almost coincide. We already observed
this twice: In \Cref{fig:veto_NLO_mH_rel} around  $\mhiggs=125$\,GeV and in
\Cref{fig:onejetLOrel} for $\ptveto\lesssim 90$\,GeV. In both cases, the dashed
curve (contributions up to $1/m_t^2$) and the dash-dotted curve (contributions
up to $1/m_t^4$) are basically on top of each other and approximate the exact
result extremely well ($<1$\%). 

Overall, our observations so far are encouraging to study the behavior of the
$1/\mtop$ expansion at higher orders to estimate the range of applicability of
the heavy-top limit for jet observables.

\subsection{Jet-veto at \nnlo{}}\label{sec:njet}

\begin{figure}
\begin{subfigure}[b]{0.5\textwidth}
    \includegraphics[width=\textwidth]{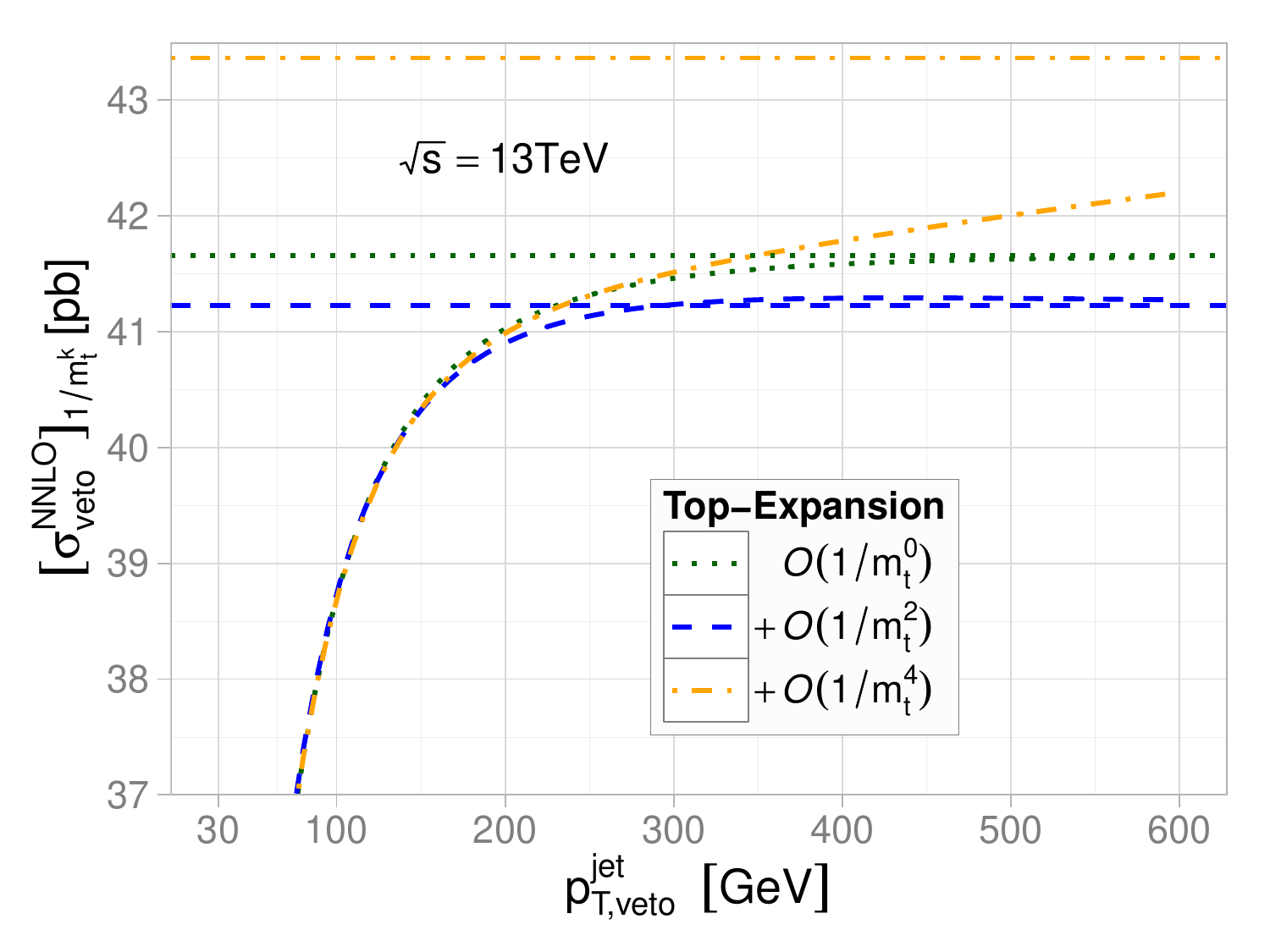}
    \caption{}
    \label{fig:jetvetoNNLO_abs}
\end{subfigure}
\begin{subfigure}[b]{0.5\textwidth}
    \includegraphics[width=\textwidth]{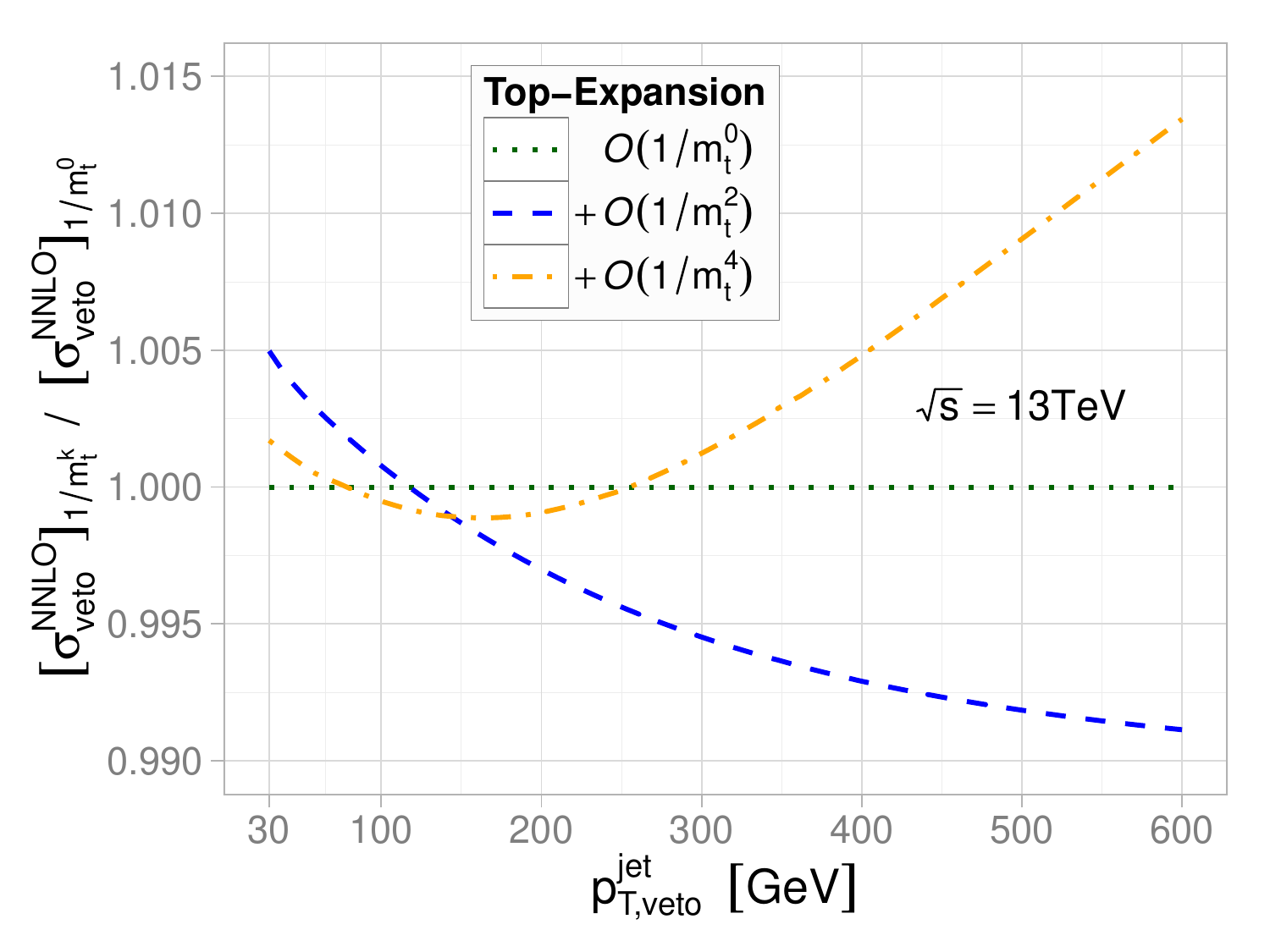}
    \caption{}
    \label{fig:jetvetoNNLO_rel}
\end{subfigure}
\begin{center}
    	\parbox{.9\textwidth}{%
      \caption{\label{fig:jetvetoNNLO}Higgs+$0$-jet cross section at \nnlo{}
      including terms up to $1/\mtop^k$ as a function of $\ptveto{}$.
  Dotted/dashed/dash-dotted: $k = 0/2/4$.}} \end{center}
\end{figure}

We are now ready to analyze the mass effects on the jet-vetoed rate at \nnlo{},
which is the central observable of our study.  \Cref{fig:jetvetoNNLO_abs} shows
the truncation of the cross section with a jet-veto at $1/\mtop^k$ for $k=0$
(dotted), $k=2$ (dashed) and $k=4$ (dash-dotted) as a function of the jet-veto
cut. At small values of $\ptveto$, we observe an excellent convergence of the 
asymptotic expansion, i.\,e. the cross section is almost independent of
the order of expansion in $1/\mtop{}$. For example, the spread of the curves is
about $0.5$\% at $\ptveto=30$\,GeV, see \Cref{fig:jetvetoNNLO_rel}, where all
curves are normalized to the \eft{} ($k=0$). In fact, $[\sigma_{\text{veto}}^\nnlo{}]_{1/\mtop^k}$ 
behaves even better with increasing $k$ than the total inclusive cross section, where a matching
to the high-energy limit is required \cite{Harlander:2009my} to alleviate the unjustified large 
effects from hard jets. These 
effects do not appear in case of the jet-vetoed cross section. More precisely, they explicitly cancel between
$\sigma^\nnlo_{\text{tot}}$ and $\sigma^{\nlo'}_{\ge 1\text{-jet}}$ in \eqn{eq:jetveto}.

At larger values of the jet-veto cut, the deviation
between the curves in \fig{fig:jetvetoNNLO} increases.  They stay remarkably 
small though ($\sim 2$\% at
$\ptveto=600$\,GeV). Thus, similarly to what we found at \nlo{}, the asymptotic
expansion of the cross section is well behaved even for a jet-veto beyond the $2\,\mtop$
threshold, because contributions from large-\pt{} jets are suppressed by
phase-space.\footnote{As we see in \sct{sec:hardest}, mass effects become
important once the transverse momenta of the hardest jet exceeds $\sim
150$\,GeV.}

\begin{figure}
\begin{center}
\includegraphics[width=0.65\textwidth]{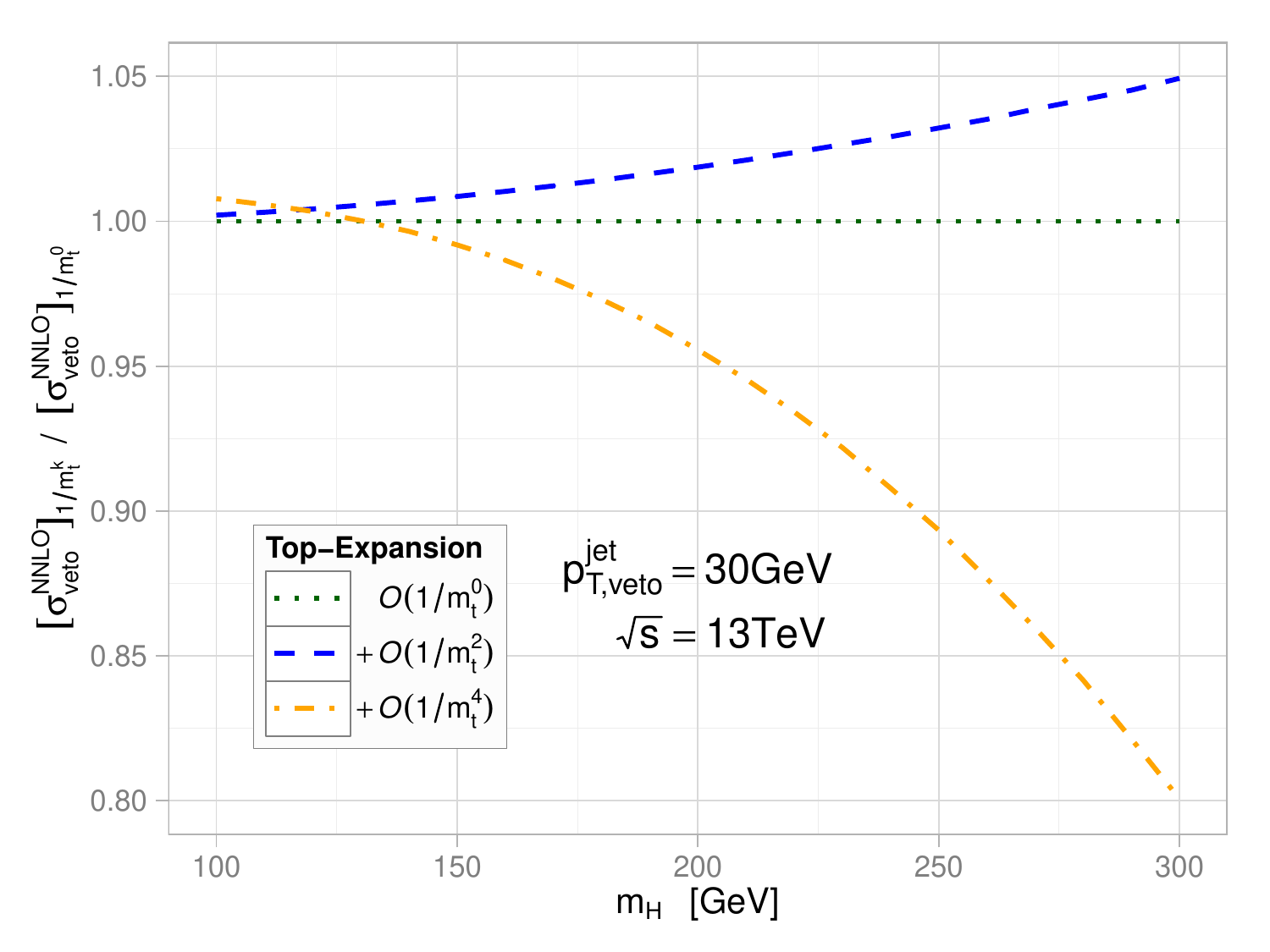}
	\parbox{.9\textwidth}{%
      \caption[]{\label{fig:mH}{Higgs+$0$-jet cross section at \nnlo{}
      including terms up to $1/\mtop^k$ as a function of $\mhiggs$
      normalized to heavy-top limit ($k=0$) for $\ptveto=30$\,GeV.
      Dotted/dashed/dash-dotted: $k = 0/2/4$.  }}}
\end{center}
\end{figure}

In a large number of beyond standard model (\bsm) theories, additional scalar
particles are predicted, e.\,g. a second (heavier) \cp{}-even Higgs boson.
Therefore, we investigate the quality of the
$\mtop\rightarrow\infty$ approximation for more general Higgs masses.
\Cref{fig:mH} shows the $1/\mtop^k$ expansion ($k=0/2/4$) of the jet-vetoed
\nnlo{} cross section normalized to the \eft{} result ($k=0$) as a function of
$\mhiggs$. Indeed, the effective field theory yields a valid approximation at
the one-percent level for $\mhiggs\lesssim 150$\,GeV. At larger Higgs masses
the top-mass effects become sizable and the uncertainty induced by the
heavy-top limit increases to $\sim 6\,(25)$\% at $\mhiggs=200\,(300)\,\mathrm{GeV}$

In summary, for a \sm{} Higgs boson of mass $125.6$\,GeV it is fully justified
to trust the effective field theory approach to determine radiative corrections
to the jet-vetoed cross section at \nnlo{}. It is advisable though
    to account for the full mass dependence at \lo{} through reweighting, as it
is common practice and done in our analysis.
Furthermore, our results should directly generalize to the resummed
jet-vetoed cross section at \nnlo{}\plus{}\nnll{} \cite{Banfi:2012jm} evaluated
in the \eft{}, since the resummation of Sudakov logarithms from soft-gluon
emissions is predominantly described by process independent \qcd{} effects.

\subsection{Inclusive Higgs+jet rate at \bld{\nlo{}}}\label{sec:incl}

\begin{figure}
\begin{subfigure}[b]{0.49\textwidth}
    \includegraphics[width=\textwidth]{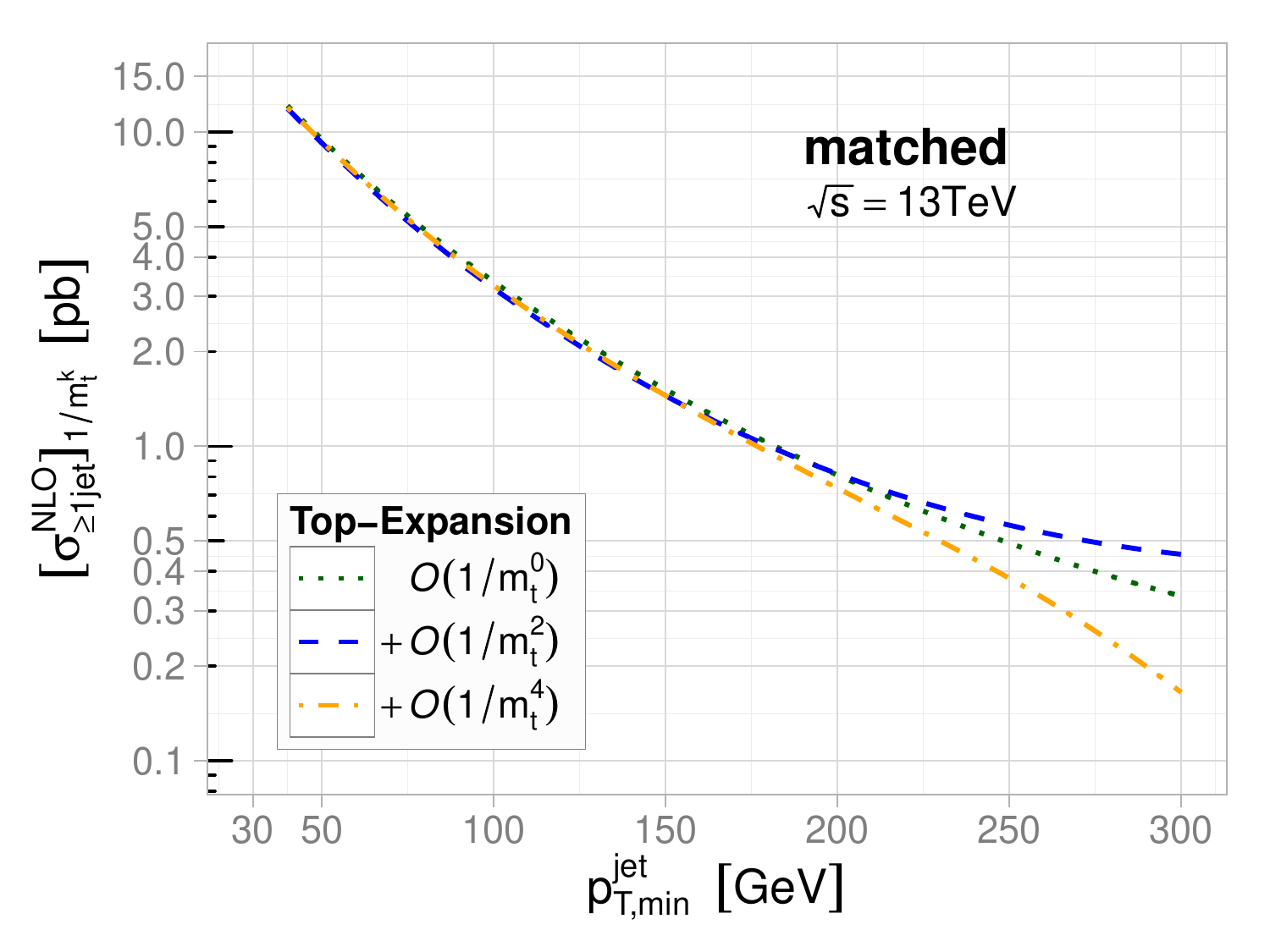}
    \caption{}
    \label{fig:onejet_matched}
\end{subfigure}
\begin{subfigure}[b]{0.49\textwidth}
    \includegraphics[width=\textwidth]{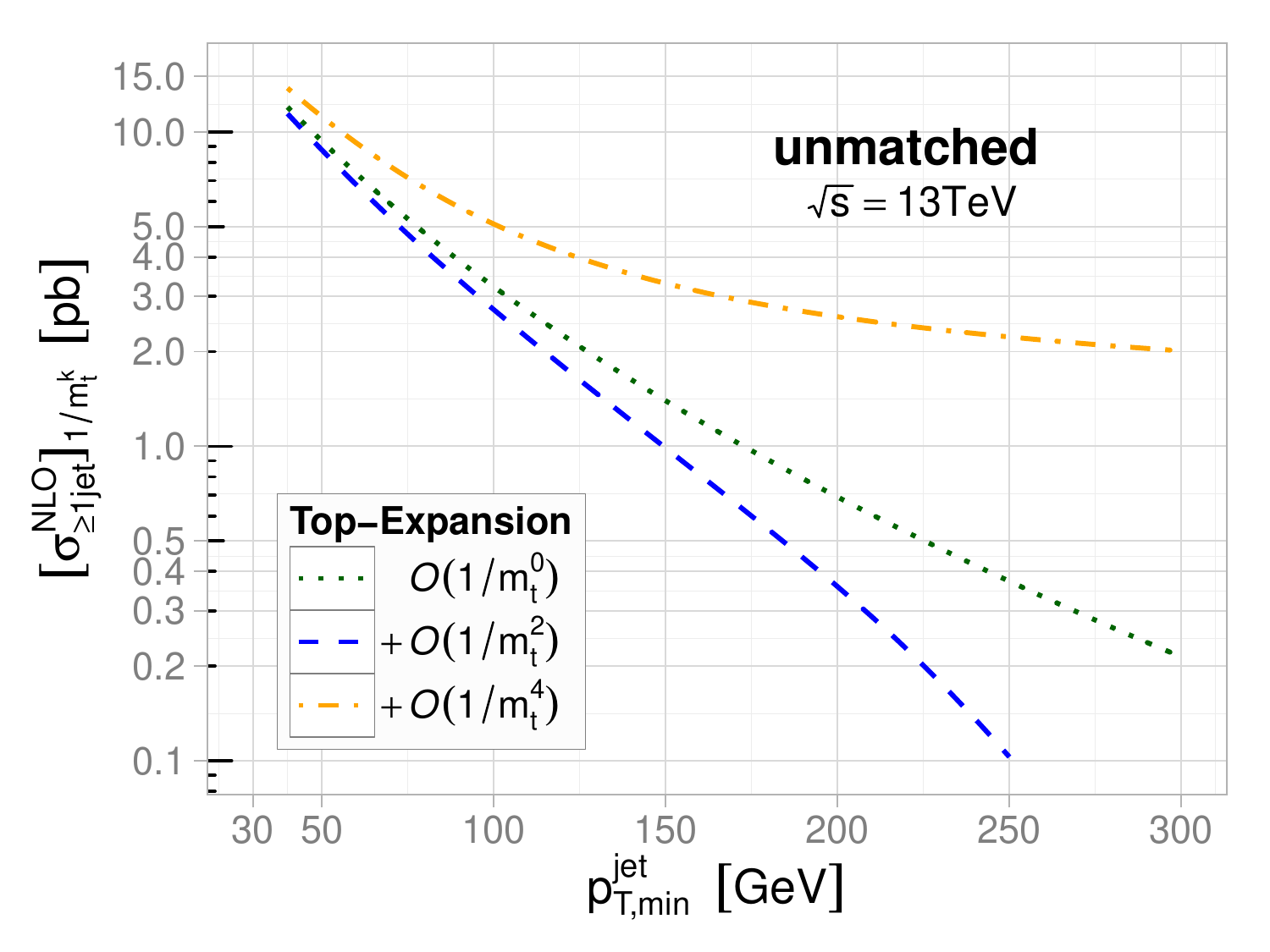}
    \caption{}
    \label{fig:onejet_unmatched}
\end{subfigure}
\begin{center}
\parbox{.9\textwidth}{%
      \caption[]{\label{fig:onejet}{Inclusive Higgs+jet cross section at \nlo{}
    including terms up to $1/\mtop^k$ as a function of $\ptmin{}$.
    Dotted/dashed/dash-dotted: $k = 0/2/4$. (a) matched
    according to \eqn{eq:match}; (b) unmatched.}}} 
\end{center}
\end{figure}

\begin{figure}
\begin{center}
    \includegraphics[width=0.65\textwidth]{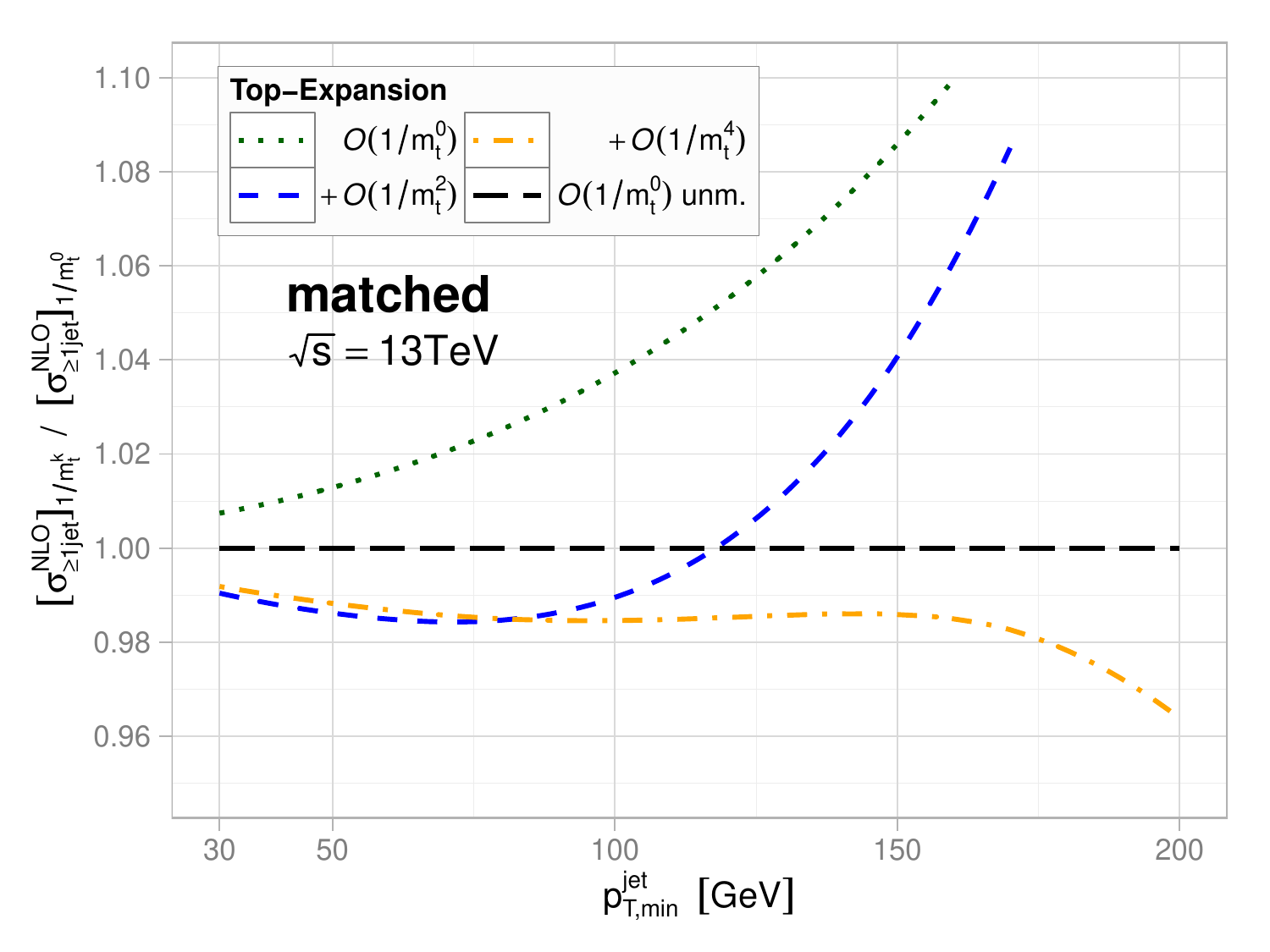}
	\parbox{.9\textwidth}{%
    \caption{\label{fig:onejet_matched_rel}Same as \Cref{fig:onejet_matched}, but
    normalized to unmatched $1/m_t^0$ cross section (dotted curve of
    \Cref{fig:onejet_unmatched}).}}
\end{center}
\end{figure}

For the \lo{} Higgs+jet cross section, the $1/\mtop$ expansion provides no
proper approximation of the top-mass effects, as we have seen in
\sct{sec:loword}. The reason for this are unjustified large contributions from
high-\pt{} jets at higher orders in $1/\mtop{}$. In order to obtain a reliable
estimate of the mass effects on the \lo{} Higgs+jet rate, we defined the
matched cross section in \eqn{eq:matchLO}. Moving to $\als^4$, we encounter
the same problems, which can be seen from the dash-dotted curve (expansion up
to $1/\mtop^4$) in \Cref{fig:jetvetoNNLO_abs} at $\ptjet\gtrsim 400$\,GeV, for
example. Consequently, not only the Higgs+jet cross section at \lo{} is
affected, but also at \nlo{}. This is why we define the matched inclusive
Higgs+jet rate at \nlo{} accordingly:
\bal
\label{eq:match}
\left[\sigma^{\nlo}_{\ge 1\text{-jet,\,matched}}\right]_{\mtop^k} \equiv \left[\sigma^{\nlo}_{\ge 1\text{-jet,\,unmatched}}\right]_{\mtop^k}  + \left[\sigma^{\nnlo^*}_{\text{tot,\,matched}}\right]_{\mtop^k}  - \left[\sigma^{\nnlo^*}_{\text{tot,\,unmatched}}\right]_{\mtop^k}  \,,
\eal
where the starred \nnlo{} cross section is calculated with \nlo{} \pdf{}s. 

The matched cross section expanded up to different orders in $1/\mtop^k$ is
shown in \Cref{fig:onejet_matched} ($k=0/2/4$). All three curves are very close,
extending the validity of the asymptotic expansion to significantly larger
values of $\ptmin$ than in the unmatched case, see \Cref{fig:onejet_unmatched}. 
\Cref{fig:onejet_matched_rel} shows the improved
matched predictions of \Cref{fig:onejet_matched} normalized to unmatched cross
section in the heavy-top limit (dotted curve of \Cref{fig:onejet_unmatched}). 
The $1/\mtop^4$ term yields a very small
correction for $\ptmin\in[30,100]$\,GeV.  In this case, we trust the dashed
(expansion up to $1/\mtop^2$) and dashed-dotted curve (expansion up to
$1/\mtop^4$) to approximate the exact mass effects to better than one percent.
Therefore, as long as the minimum jet-\pt{} cut remains at moderate values ($\ptmin\lesssim 100$\,GeV) 
the definition of the matched cross section in \eqn{eq:matchLO} and \eqn{eq:match} 
allows us to determine a reliable prediction of the inclusive Higgs+jet rate at \lo{} and
\nlo{}, respectively. Furthermore, comparing the matched curve at $1/\mtop^4$ 
to the unmatched \eft{} 
result, we validate the heavy-top approximation at the level of $1$-$2$\% for 
$\ptmin\le 100$\,GeV.

This result shows that the \eft{}, in fact, works better in the problematic
high-\pt{} region than the corresponding sub-leading $1/\mtop$ terms, which
are far apart in the unmatched case, see \Cref{fig:onejet_unmatched}. This is very similar to what was found
for the total cross section \cite{Harlander:2009my}, where it was argued that
in the heavy-top limit ($k=0$) problematic terms $(\sqrt{s}/\mtop)^k$ vanish,
which spoil the convergence of the asymptotic expansion ($k>0$) in the
high-energy region. Also in this case the matching to the high-energy limit
revealed that the unmatched \eft{} result is valid at the percent level.

 However, at larger values ($\ptmin> 100$\,GeV), the asymptotic expansion starts deteriorating 
significantly already for the matched cross section in \fig{fig:onejet_matched_rel}. Therefore, the uncertainty 
of the \eft{} due to mass effects in that region is quite sizable. For comparison,
the deviation of the \eft{} from the exact curve is $12$($30$)\% for 
$\ptmin=100(200)$\,GeV at \lo{}, see \fig{fig:onejetLOrel}.

\subsection{Distributions of the hardest jet}\label{sec:hardest}

\begin{figure}
\begin{center}
\includegraphics[width=0.8\textwidth]{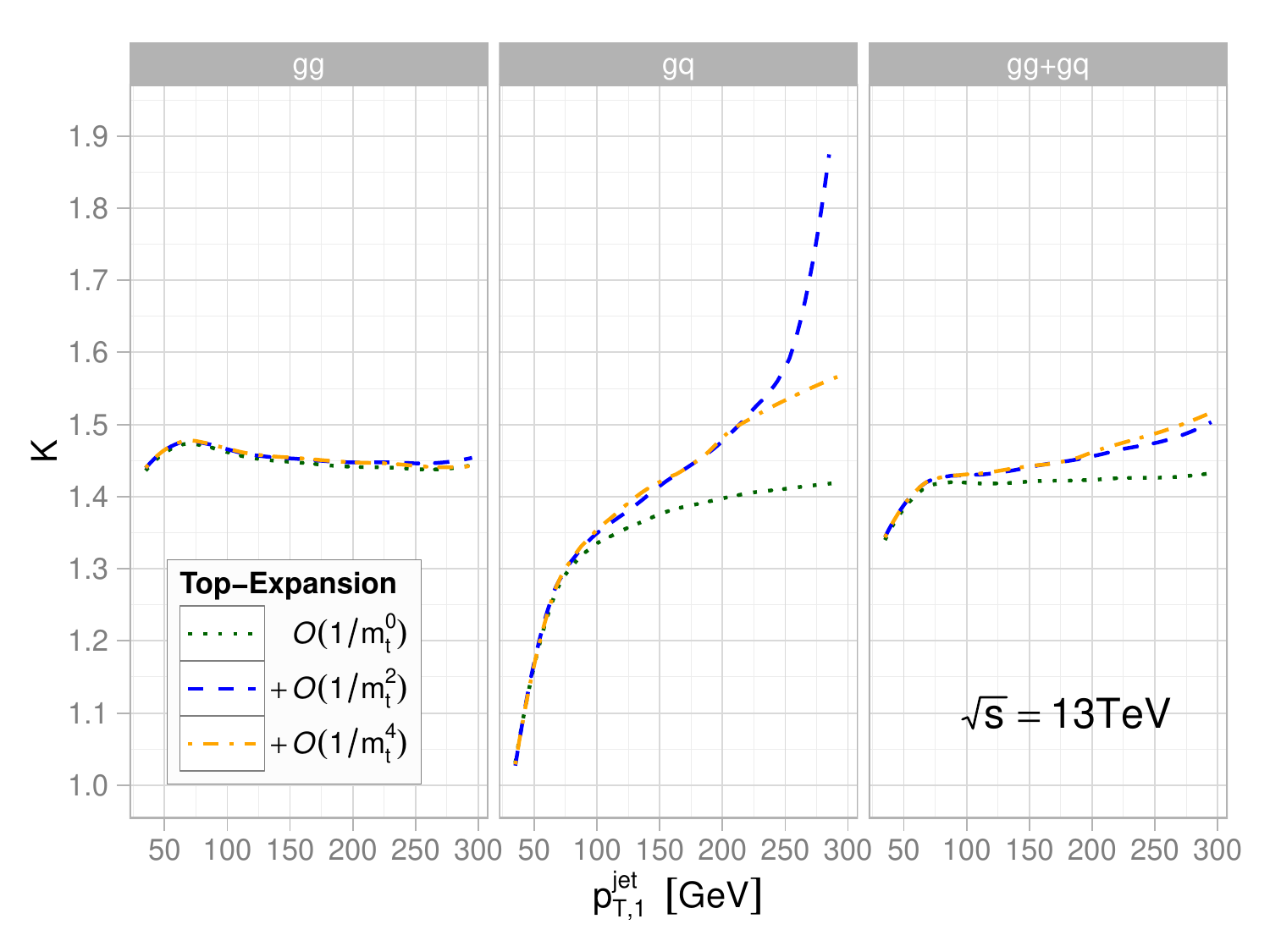}
\parbox{.9\textwidth}{
    \caption{\label{fig:pTj}$K$-factors as defined in \eqn{eq:K}, for the
    transverse momentum distribution of the hardest jet, i.\,e. $K_k^\nlo{} \equiv
    K_k^\nlo{}(\ptjetone)$. Left/center/right plot: only $gg$/only $qg$/sum of $gg$
    and $qg$. Dotted/dashed/dash-dotted: $k = 0/2/4$.  }}
\end{center}
\end{figure}

Finally, let us consider kinematical distributions of the hardest jet.
\Cref{fig:pTj} shows the  \pt{}-dependent $K$-factors $K_k^\nlo{}\equiv
K_k^\nlo{}(\ptjetone)$ of the cross section up to $1/\mtop^k$ as defined in
\eqn{eq:K} with variable scales
\bal
\muF=\muR=\sqrt{m_H^2+(\ptjetone)^2}\,.
\eal
 In the $gg$-channel, all three $K$-factors are almost identical.
However, the \qcd{} corrections to the subleading mass terms in the
$qg$-channel behave quite differently to the \eft{} result once $\ptjetone
\gtrsim 100$\,GeV. In the sum of both channels though, the difference remains
below $\sim 1.5$\% for $\ptjetone < 150$\,GeV, and reaches $6$\% at
$\ptjetone = 300$\,GeV. Therefore, our results turn out to be
quite similar to what was already found for the \pt{} distribution of the Higgs
$K_k^\nlo{}(p_T^H)$ \cite{Harlander:2012hf}, yet the asymptotic behavior is
slightly improved for the hardest jet. For comparison, we give an updated
result for $K_k^\nlo{}(p_T^H)$ up to $1/\mtop^4$ in \app{app:pTH}, which shows
that $K_4^\nlo{}$ behaves quite differently in the two cases at high \pt{}.

Note that the $\mtop^4$ corrections are extremely small for $\ptjetone\lesssim 200$\,GeV
in \Cref{fig:pTj}.  We conclude therefore that the quality of 
$K_2^\nlo{}$ and $K_4^\nlo{}$ to approximate the exact top-mass effects 
is better that one percent in that region.

\begin{figure}
\begin{center}
\hspace{-2cm}
\includegraphics[width=0.8\textwidth]{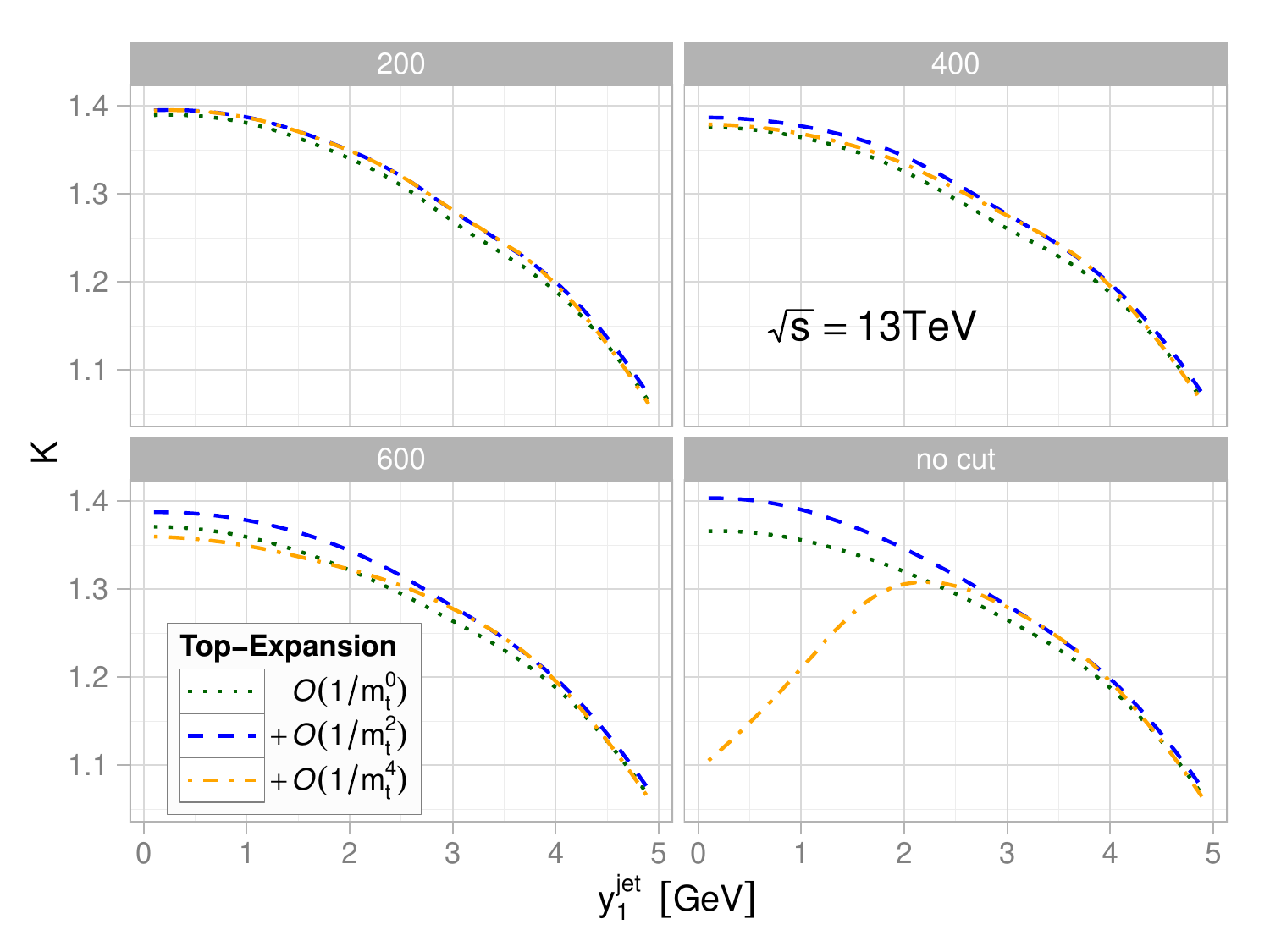}
\end{center}
    \parbox{.9\textwidth}{
       \caption{\label{fig:yj}$K$-factors as defined in \eqn{eq:K}, for the
       rapidity distribution of the hardest jet, i.\,e. $K_k^\nlo{}\equiv
       K_k^\nlo{}(\yjetone)$. Left-top/right-top/left-bottom/right-bottom plot:
       $\ptmax=200$\,GeV/$400$\,GeV/$600$\,GeV/no cut.
       Dotted/dashed/dash-dotted: $k = 0/2/4$.
        }}
\end{figure}

The situation for the rapidity distribution of the hardest jet is more involved.
The problem is that in the central region the $1/\mtop^4$ term receives
unjustified large effects from hard jets, which spoil the convergence of the
asymptotic series. Unfortunately, it is not possible to determine a matched
cross section in this case, similarly to what we do for the inclusive
Higgs+jet cross section. Instead, we introduce a cut $\ptjet<\ptmax$ which
simply removes the problematic high-\pt{} jets. This cut is of course
arbitrary, therefore, we choose three different values:
$\ptmax=200,\,400,\,600$\,GeV. 

In fact, the contribution to the $\yjetone$
distribution from jets with $\ptjet>600$\,GeV should be completely negligible
due to phase-space suppression. This is what we observe for the \eft{} result,
but not for the subleading terms in the $1/\mtop$ expansion, see \Cref{fig:yj},
which shows $K_k^\nlo{}\equiv K_k^\nlo{}(\yjetone)$ for
$\ptmax=200,\,400,\,600$\,GeV and without cut. Clearly, the
asymptotic behavior in the central region is broken without a cut. It works
pretty well though once we apply an upper cut on the jets. 
The \eft{} result is almost identical ($<0.5$\%) in the lower two plots and receives no
unjustified large effects from high-\pt{} jets. 
Therefore, it is legitimate to estimate the quality of the \eft{} without a cut
from the results for $\ptmax=600$\,GeV, which we deduce to be better than
$2$\% in the central region and even below one percent in
the forward region ($\yjetone>2.5$). 

In conclusion, the behavior of the $K$-factors of the hardest jet
distributions suggest that the \qcd{} corrections can be safely calculated in
the heavy-top approximation. The accuracy remains within 1.5\% (6\%) below
$\ptjetone =150$\,GeV ($\ptjetone = 300$\,GeV) and for \pt{}-integrated
quantities at the percent level.

\section{Conclusions}\label{sec:conclusions}

Finite top-mass effects in the gluon fusion process have been studied. The quality of the effective field theory to describe the exact cross section was estimated using subleading terms in $1/\mtop$. They have been evaluated for various jet quantities, namely, the \nnlo{} cross section with a jet veto, the inclusive Higgs+jet rate at \nlo{} and the \nlo{} $K$-factors of jet distributions.

The corrections of a finite top-mass to the jet-vetoed rate are negligible and the quality of the effective field theory to describe this quantity even at large values of the jet-veto cut is remarkable. Unjustified large contribution from hard jets were found to spoil the convergence of the asymptotic expansion in case of the inclusive Higgs+jet cross section. Only a matching procedure involving the total inclusive cross section allowed for a reliable prediction of this quantity and the estimation of the mass effects from the $1/\mtop{}$ expansion. The \eft{} was then found to be valid even without the matching at the $1$-$2$\% level for jet definitions with a minimal transverse momentum cut lower than $100$\,GeV.
For large values though, the asymptotic expansion of the matched result becomes unreliable. Therefore, also the uncertainty induced by the \eft{} is large, deviating by $30$\% from the exact result already at LO for a minimal jet cut of $200$\,GeV.

Also the perturbative corrections to distributions of the hardest jet turned out to have a rather mild top-mass dependence. For the transverse momentum distribution, the procedure of correcting the \lo{} prediction including the exact top-mass dependence by the $K$-factor evaluated in the \eft{} provides an excellent approximation to the full \nlo{} result, valid to better than $1.5$(5)\% for $\ptjet<150$(300)\,GeV. The $K$-factor of the rapidity distribution determined in the heavy-top limit was validated at the $1$-$2$\% level.

We have checked that our results hold also for different machine energies at the \lhc{}. The accuracy of the
effective field theory approach is better than the uncertainty on the cross section induced by the PDFs and missing higher order QCD corrections.

\paragraph{Acknowledgements.}
We would like to thank Pier Francesco Monni for useful comments on the manuscript. We are indebted to Robert Harlander for fruitful discussions, his comments on the manuscript and the private version of
his code {\tt ggh@nnlo} that he provided for our study. The work of TN was
supported by {\abbrev BMBF} contract 05H12PXE. MW was supported by the European Commission through the FP7 Marie Curie Initial Training Network ``LHCPhenoNet'' (PITN-GA-2010-264564).

\appendix \gdef\thesection{Appendix \Alph{section}}

\section{Higgs \bld{\pt{}} distribution}\label{app:pTH}

\begin{figure}[h]
\begin{center}
\includegraphics[width=0.8\textwidth]{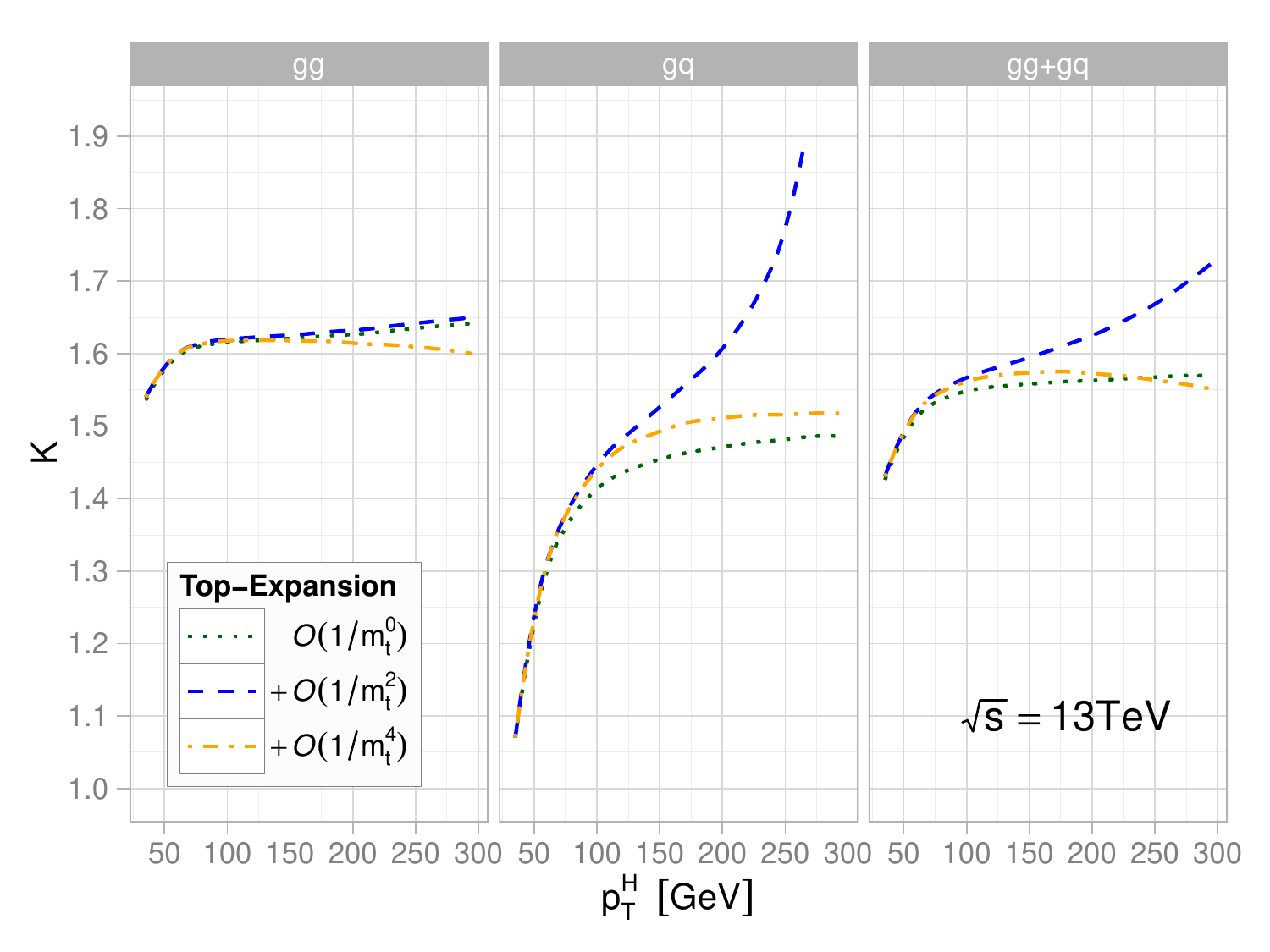}
    \parbox{.9\textwidth}{
      \caption[]{\label{fig:pTH}{Same as \Cref{fig:pTj}, but for the transverse
      momentum distribution of the Higgs; here: $K_k^\nlo{} \equiv
      K_k^\nlo{}({p_T^H})$.  } }}
\end{center}
\end{figure}

For completion, we update the results of \citere{Harlander:2012hf} for the $K$-factors of the transverse momentum distribution of the Higgs $K_k^\nlo{}(p_T^H)$, see \fig{fig:pTH}. The factorization and renormalization scale are set to the transverse mass of the Higgs 
\bal
\muF=\muR=m_T^H=\sqrt{m_H^2+(\pt^H)^2}\,.
\eal
Additionally to \citere{Harlander:2012hf}, we determine the $K$-factor expanded up to $1/\mtop^4$. Our result perfectly confirms the conclusions drawn in that paper, since $K_4$ lies just right between $K_0$ and $K_2$ for most transverse momenta.

\bibliographystyle{jetveto}
\bibliography{jetveto}

\end{document}